# Dispersive wave propagation in two-dimensional rigid periodic blocky materials with elastic interfaces

Andrea Bacigalupo[1] and Luigi Gambarotta[2*]

[1]IMT School for Advanced Studies, Lucca, Italy
[2]Department of Civil, Chemical and Environmental Engineering, University of Genova, Italy

**Abstract**

Dispersive waves in two-dimensional blocky materials with periodic microstructure made up of equal rigid units having polygonal centro-symmetric shape with mass and gyroscopic inertia, connected each other through homogeneous linear interfaces, have been analysed. The acoustic behavior of the resulting discrete Lagrangian model has been obtained through a Floquet-Bloch approach. From the resulting eigenproblem derived by the Euler-Lagrange equations for harmonic wave propagation, two acoustic branches and an optical branch are obtained in the frequency spectrum. A micropolar continuum model to approximate the Lagrangian model has been derived based on a second-order Taylor expansion of the generalized macro-displacement field. The constitutive equations of the equivalent micropolar continuum have been obtained, with the peculiarity that the positive definiteness of the second-order symmetric tensor associated to the curvature vector is not guaranteed and depends both on the ratio between the local tangent and normal stiffness and on the block shape. The same results has been obtained through an extended Hamiltonian derivation of the equations of motion for the equivalent continuum that is related to the Hill-Mandel macro homogeneity condition. Moreover, it is shown that the hermitian matrix governing the eigenproblem of harmonic wave propagation in the micropolar model is exact up to the second order in the norm of the wave vector with respect to the same matrix from the discrete model. To appreciate the acoustic behavior of some relevant blocky materials and to understand the reliability and the validity limits of the micropolar continuum model, some blocky patterns have been analysed: rhombic and hexagonal assemblages and running bond masonry. From the results obtained in the examples, it appears that the obtained micropolar model turns out to be particularly accurate to describe dispersive functions for wavelengths greater than 3-4 times the characteristic dimension of the block. Finally, in consideration that the positive definiteness of the second order elastic tensor of the micropolar model is not guaranteed, the hyperbolicity of the equation of motion has been investigated by considering the Legendre–Hadamard ellipticity conditions requiring real values for the wave velocity.

---

[*] Corresponding Author, luigi.gambarotta@unige.it

## 1. Introduction

Periodic material microstructures having a rigid phase with dominant volumetric fraction and a soft phase with a vanishing volume fraction are common in blocky rock systems, granular materials, masonry, masonry-like biological and nacreous bio-inspired heterogeneous composites. Recent studies on nacreous-like materials, having a microstructure similar to the running bond masonry, have shown interesting phononic properties of these materials, in particular the existence of band gaps (Chen and Wang, 2014, 2015, Yin *et al.*, 2014) and the possibility to obtain devices for vibration reduction and isolation (Yin *et al.*, 2015). Although these studies concerning the wave propagation have been carried out considering the material as a periodic continuum and by applying the finite element Bloch approach, approximate models of blocky materials can be useful to control and optimize the acoustic performance and to check results provided by detailed models.

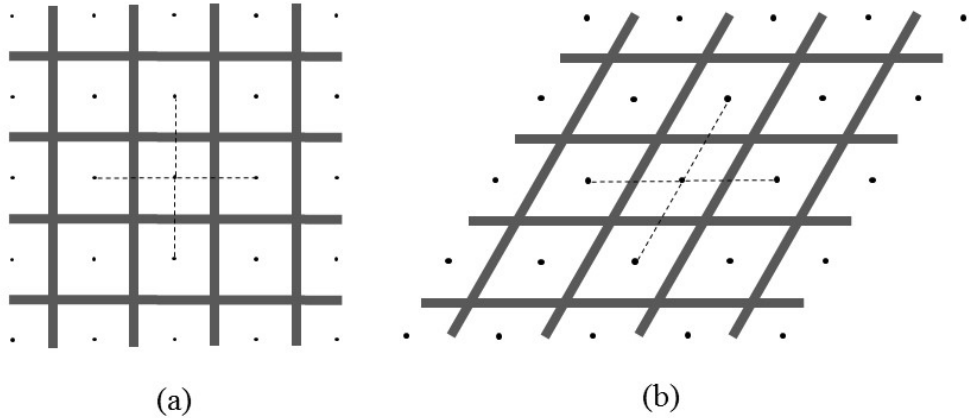

Figure 1: Examples of periodic centro-symmetric rigid block tessellation with elastic interfaces (in dark grey) (coordination number *n*=4).

An approximate description of these materials can be obtained by assuming the dominant phase as a rigid body and by locating the deformation in the soft interfaces. If the interfaces between the blocks are modeled as elastic with vanishing thickness, the plane blocky system may be described by a Lagrangian model in which each block is equipped with in-plane translation and rotation. Moreover, if the characteristic size of the blocks is negligible compared to the whole body size, then the discrete system may be approximated by a homogeneous equivalent continuum. However, for this class of composites the classical (Cauchy) homogenization may have disadvantages and non-local constitutive models may be necessary to include geometric and material length scales to appreciate the influence of block size and of high stress and strain gradients (see for instance Bacigalupo and Gambarotta, 2011, 2012, 2014, and Bacigalupo, 2014). The rotational degrees of freedom of the blocks suggests to consider the Cosserat model as equivalent continuum and accordingly homogenization techniques in the static field have been proposed for blocky rocks (Mühlhaus, 1993), granular regularly packed materials (see Kruyt, 2003, Li *et al.*, 2010, among the others) and periodic masonry (Masiani *et al.*, 1995, Trovalusci and Masiani, 2003, 2005, Pau and Trovalusci, 2012, Baraldi *et al.*, 2015, among the others).

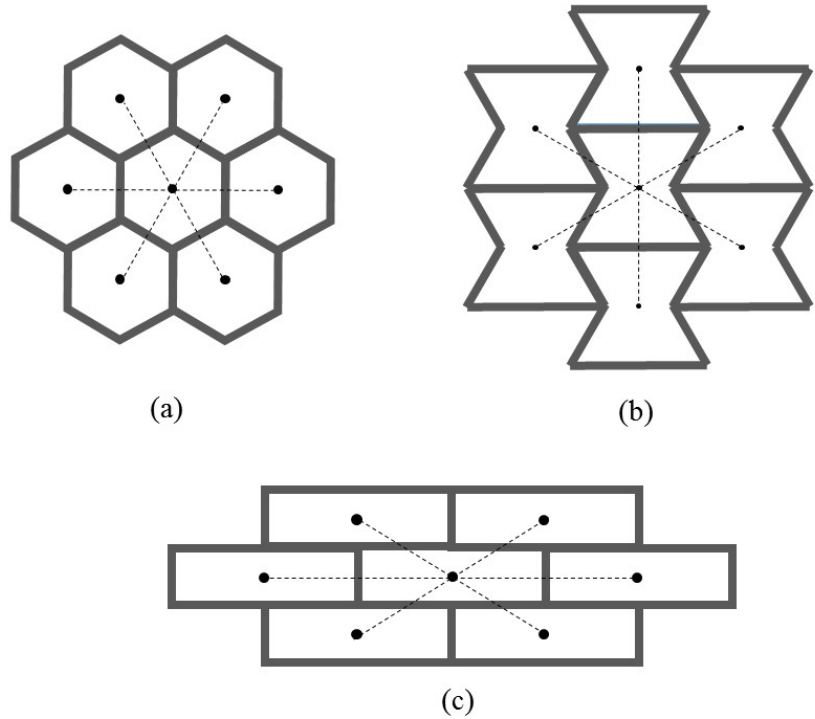

Figure 2: Examples of periodic centro-symmetric rigid block tessellation with elastic interfaces (coordination number *n*=6).

Dispersive propagation of harmonic waves in discrete blocky systems have been analysed by Sulem and Mühlhaus, 1997, who complemented their results with the identification of a Cosserat continuum in order to approximate the band structure (i.e. the dispersive functions) of the discrete model. A similar analysis was carried out for granular materials by Suiker *et al.*, 2001, and Suiker and de Borst, 2005, where a second order micropolar continuum was also considered, and its results compared to those from the Lagrangian model. The wave propagation in running bond masonry was analyzed by Stefanou *et al.*, 2008, where the bricks were assumed as rigid bodies. The validity limits of the Cosserat model were derived from the comparison of the dispersion functions with those from the discrete model. The same Authors (Stefanou *et al.*, 2010) extended this study to the case of diatomic masonry patterns and proposed an equivalent micromorphic continuum to approximate the discrete blocky system. More recently, Sadovskii and Sadovskaya, 2010, and Sadovskaya and Sadovskii, 2015, analyzed the propagation of waves in blocky materials modeled as an equivalent orthotropic Cosserat continuum. Finally, it is worth noting that Merkel *et al.*, 2011, have highlighted some limitations of the Cosserat model in simulating the optical band structure of hexagonally packed granular materials in the neighborhood of long waves. In particular, these Authors criticized the inability of the Cosserat model to simulate the downward concavity of the optical branch in the Bloch domain obtained in the Lagrangian model for long rotational wavelengths. This discrepancy between the micropolar and the discrete model may be observed also in the results obtained on running bond and diatomic masonry by Stefanou *et al.*, 2008, 2010.

The present paper is focused on simplified modeling of periodic blocky materials with centro-symmetric blocks connected by straight elastic interfaces having evanescent thickness. The blocks are assumed to be rigid bodies and a discrete model is formulated from which the equations governing the propagation of harmonic waves are derived. The influence of the model parameters on the dispersion functions, providing the wave angular frequency as a function of the wave vector in the Brilluoin domain are analyzed. By applying a continualization technique, which provides the generalized block displacements via a second order expansion of the macro-displacement field, the elastic tensors of the equivalent Cosserat continuum are identified. In particular, it comes out that the elastic tensor which couples the curvatures to the micro-couples may be negative defined. This result is also obtained from an application of the generalized Hill-Mandel macro-homogeneity condition. The equations of wave propagation in the equivalent Cosserat continuum are derived for the rather general case of centrosymmetric blocks.

Three examples of blocky materials are examined: rhombic and hexagonal pattern and finally running bond masonry. For each assemblages, the equations of wave propagation in the discrete model and in the micropolar continuum are inferred, along with the elastic constants. Some numerical examples are analyzed to appreciate both the influence of the model parameters on the band structure and to assess the validity limits of the Cosserat continuum. This investigation is also aimed at understanding and solving the mentioned criticism by Merkel *et al.*, 2011, regarding the capability of the Cosserat model to approximate the optical branch. Although the present formulation provides a constitutive model that in several cases is characterized by negative definiteness of the elastic tensor of the microcurvatures, however, the existence and uniqueness of the solution of the dynamic problems considered in the examples has been verified through the ellipticity condition of Legendre-Hadamard, namely by verifying that the wave phase velocity in the elastic micropolar continuum is real.

## 2. Dispersive waves in periodic blocky materials with elastic interfaces

The blocky materials considered in the previous Section (see Figures 1 and 2) are modeled as 2-D assemblages made up of equal rigid blocks having polygonal centro-symmetric shape, periodically arranged in the plane and connected each other through homogeneous linear elastic interfaces. The topology of the blocky materials is limited to the more common case of even coordination number *n*, i.e. the number of contacts that each block has with other surrounding blocks (see the examples in Figure 1 with *n*=4 and in Figure 2 with *n*=6). Let consider now a generic reference block surrounded by *n* of blocks through interfaces. The gravity center of the *i*-th surrounding block is identified by vector $\mathbf{x}_i = l_i \mathbf{n}_i$ with respect to the gravity center of the reference block, $l_i$ and $\mathbf{n}_i$ being the distance between the two block centers and the unit vector related to the *i*-th block, respectively (the unit vector $\mathbf{t}_i$ in Figure 3.a is normal to $\mathbf{n}_i$). The interface is modelled as a linear element of length $b_i$ and vanishing thickness, its orientation being represented by the unit vector $\mathbf{d}_i$ normal to the interface as shown in figure 3(a). Because of the assumption of centro-symmetry of the blocks, the analysis is limited to the cases in which the interface is located at the mid-point of vector $\mathbf{x}_i$.

The in-plane rigid motion of the blocks is represented by the translation and rotation of their (gravity) centers. As shown in figure 3(b) the motion of the reference block is described by $(\mathbf{u}, \phi)$, and for the *i*-th block by $(\mathbf{u}_i, \phi_i)$. Each block has mass M and mass moment of inertia $J = Mr^2$. The relative displacement between two contiguous blocks is represented by the relative displacement $\Delta \mathbf{s}_i$ and the relative rotation $\Delta \varphi_i$ at the mid-point of the *i*-th interface as shown in figure 4. Since the interface is assumed linear elastic, the local normal and tangential stiffness $k_n$ and $k_t$ are introduced, so that the elastic potential energy of the $i$-th interface between the reference and the $i$-th block is written as

$$\pi_i = \frac{1}{2} K_n^i \Delta s_{ni}^2 + \frac{1}{2} K_t^i \Delta s_{ti}^2 + \frac{1}{2} K_\varphi^i \Delta \varphi_i^2 \qquad (1)$$

having defined the overall stiffnesses $K_n^i = k_n b_i$, $K_t^i = k_t b_i$ and $K_\varphi^i = k_n b_i^3 / 12$ of the interface, respectively. Here, the normal $\Delta \mathbf{s}_{ni} = \mathbf{P}_i^\parallel \Delta \mathbf{s}_i$ and tangential $\Delta \mathbf{s}_{ti} = \mathbf{P}_i^\perp \Delta \mathbf{s}_i$ relative displacements at the interface and $\Delta \varphi_i$ the relative rotation are introduced, being $\mathbf{P}_i^\parallel = \mathbf{d}_i \otimes \mathbf{d}_i$ and $\mathbf{P}_i^\perp = \mathbf{I} - \mathbf{d}_i \otimes \mathbf{d}_i$ the projection tensors related to the unit vector $\mathbf{d}_i$.

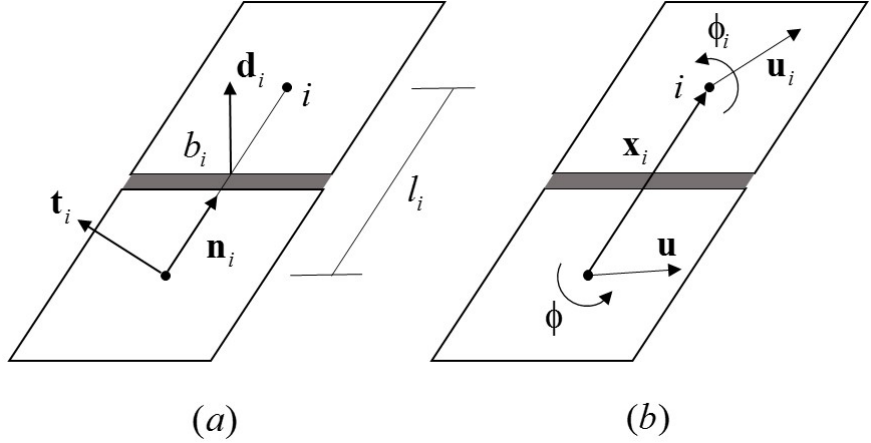

Figure 3: Two blocks connected through the interface: (a) the reference block and the adjacent $i$-th block of the $n$ blocks in contact; (b) degrees of freedom of the two blocks.

Under the above assumptions a Lagrangian system may be identified having $3N$ degrees of freedom, being $N$ the number of blocks. Under the further hypothesis of ignoring dissipative mechanisms, the Lagrangian function $\mathcal{L} = T - \Pi$ may be introduced, being $T$ the kinetic energy and $\Pi$ the elastic potential energy stored in the elastic interfaces. The kinetic energy of the reference block is written as

$$T(\dot{\mathbf{u}}, \dot{\phi}) = \frac{1}{2} M \dot{\mathbf{u}}^2 + \frac{1}{2} J \dot{\phi}^2. \qquad (2)$$

The elastic potential energy stored in the $i$-th interface is written in terms of the generalized relative displacement of the adjacent blocks in the form $\Delta \mathbf{s}_i = (\mathbf{u}_i - \mathbf{u}) - (\phi + \phi_i) \frac{l_i}{2} \mathbf{t}_i$, being $\mathbf{t}_i = \mathbf{e}_3 \times \mathbf{n}_i$ (see figure 4), so that the relative generalized displacements at the interface under consideration are

$$\Delta \mathbf{s}_{ni} = \mathbf{P}_i^{\parallel} \left[ (\mathbf{u}_i - \mathbf{u}) - (\phi + \phi_i) \frac{l_i}{2} \mathbf{t}_i \right] \quad ,$$

$$\Delta \mathbf{s}_{ti} = \mathbf{P}_i^{\perp} \left[ (\mathbf{u}_i - \mathbf{u}) - (\phi + \phi_i) \frac{l_i}{2} \mathbf{t}_i \right] \quad , \qquad (3)$$

$$\Delta \varphi_i = \phi_i - \phi \quad .$$

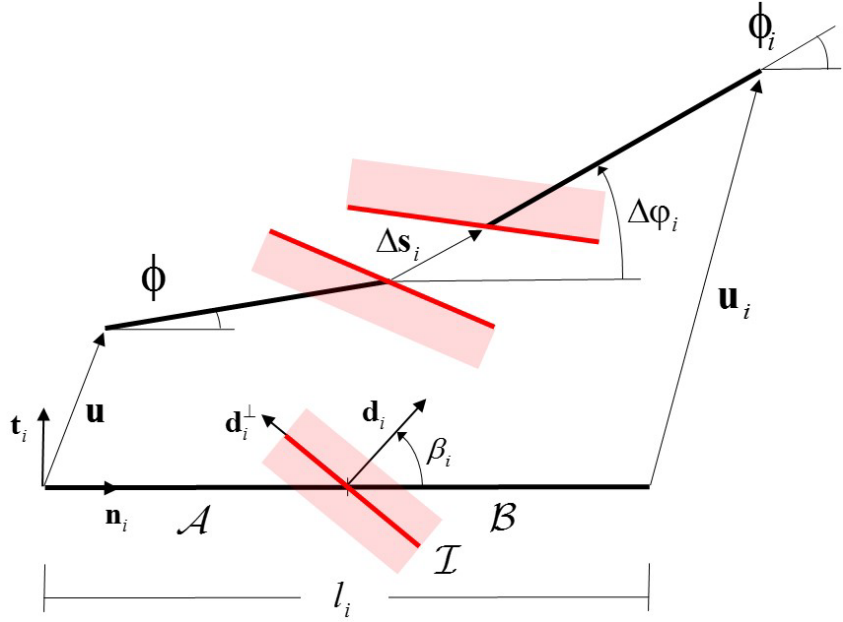

Figure 4: Relative generalized displacements at the interface $\mathcal{I}$ $I$ of the two contiguous blocks denoted by $\mathcal{A}$ and $\mathcal{B}$ respectively.

By substituting equation (3) in (1) the interface elastic energy takes the form

$$\Pi_i(\mathbf{u}, \phi, \mathbf{u}_i, \phi_i) = \frac{1}{2} K_n^i \left[ (\mathbf{u}_i - \mathbf{u}) \cdot \mathbf{P}_i^{\parallel} (\mathbf{u}_i - \mathbf{u}) + \frac{l_i^2}{4} \mathbf{t}_i \cdot \mathbf{P}_i^{\parallel} \mathbf{t}_i (\phi + \phi_i)^2 - l_i (\mathbf{u}_i - \mathbf{u}) \cdot \mathbf{P}_i^{\parallel} \mathbf{t}_i (\phi + \phi_i) \right] +$$

$$+ \frac{1}{2} K_t^i \left[ (\mathbf{u}_i - \mathbf{u}) \cdot \mathbf{P}_i^{\perp} (\mathbf{u}_i - \mathbf{u}) + \frac{l_i^2}{4} \mathbf{t}_i \cdot \mathbf{P}_i^{\perp} \mathbf{t}_i (\phi + \phi_i)^2 - l_i (\mathbf{u}_i - \mathbf{u}) \cdot \mathbf{P}_i^{\perp} \mathbf{t}_i (\phi + \phi_i) \right] + \frac{1}{2} K_{\varphi}^i (\phi_i - \phi)^2 = \qquad (4)$$

$$= \frac{1}{2} \left\{ (\mathbf{u}_i - \mathbf{u}) \cdot \mathbf{K}_i (\mathbf{u}_i - \mathbf{u}) + \frac{l_i^2}{4} \mathbf{t}_i \cdot \mathbf{K}_i \mathbf{t}_i (\phi + \phi_i)^2 - l_i (\mathbf{u}_i - \mathbf{u}) \cdot \mathbf{K}_i \mathbf{t}_i (\phi + \phi_i) \right\} + \frac{1}{2} K_{\varphi}^i (\phi_i - \phi)^2$$

having defined $\mathbf{K}_i = \left( K_n^i \mathbf{P}_i^{\parallel} + K_t^i \mathbf{P}_i^{\perp} \right)$ the symmetric second order stiffness tensor of the interface, being $\mathbf{P}_i^\| \mathbf{P}_i^\| = \mathbf{P}_i^\| = \mathbf{d}_i \otimes \mathbf{d}_i$ and $\mathbf{P}_i^\perp \mathbf{P}_i^\perp = \mathbf{P}_i^\perp = \mathbf{I} - \mathbf{d}_i \otimes \mathbf{d}_i$. Denoted with $\beta_i$ the measure of the angle between the unit vector $\mathbf{d}_i$ normal to the interface and the unit vector $\mathbf{n}_i$, the stiffness tensor may be written in the following form

$$\mathbf{K}_i = K_n^i \left( \mathbf{d}_i \otimes \mathbf{d}_i \right) + K_t^i \left( \mathbf{d}_i^\perp \otimes \mathbf{d}_i^\perp \right) =$$
$$= \left( K_n^i \cos^2 \beta_i + K_t^i \sin^2 \beta_i \right) \mathbf{n}_i \otimes \mathbf{n}_i + \left( K_n^i - K_t^i \right) \sin 2\beta_i \, \text{sym}\left( \mathbf{t}_i \otimes \mathbf{n}_i \right) + \left( K_n^i \sin^2 \beta_i + K_t^i \cos^2 \beta_i \right) \mathbf{t}_i \otimes \mathbf{t}_i.$$
(5)

The resulting Lagrangian function is obtained as the sum of the contributions of the kinetic energy of the blocks and the elastic potential energy of all the interfaces, summed on the coordination number $n$

$$\mathcal{L} = \sum \left\{ T_s \left( \dot{\mathbf{u}}, \dot{\phi} \right) - \sum_{i=1}^{n} \Pi_i \left( \mathbf{u}, \phi, \mathbf{u}_i, \phi_i \right) \right\}, \quad (6)$$

where the index denoting each block has been omitted. The Euler-Lagrange equations of motion of the reference block are obtained by the Lagrangian (6) and depend on the generalized displacement and velocity of both the reference block $(\mathbf{u}, \phi)$ and of the $n$ surrounding blocks $(\mathbf{u}_i, \phi_i)$ and are written in the following form

$$\sum_{i=1}^{n} \left[ \mathbf{K}_i \left( \mathbf{u} - \mathbf{u}_i \right) + \frac{l_i}{2} \mathbf{K}_i \mathbf{t}_i \left( \phi + \phi_i \right) \right] + \mathbf{M}\ddot{\mathbf{u}} = \mathbf{0}, \quad (7)$$

$$\sum_{i=1}^{n} \left[ \frac{l_i}{2} \mathbf{t}_i \cdot \mathbf{K}_i \left( \mathbf{u} - \mathbf{u}_i \right) + K_\varphi^i \left( \phi - \phi_i \right) + \frac{l_i^2}{4} \mathbf{t}_i \cdot \mathbf{K}_i \mathbf{t}_i \left( \phi + \phi_i \right) \right] + J\ddot{\phi} = 0, \quad (8)$$

being $\mathbf{t}_i \cdot \mathbf{K}_i \mathbf{t}_i = K_n^i \sin^2 \beta_i + K_t^i \cos^2 \beta_i$.

In case each interface is normal to the connecting vector, namely $\mathbf{d}_i = \mathbf{n}_i$, $i = 1, n$, as shown for instance in figures 1(a) and 2(a), one obtains $\mathbf{P}_i^\| = \mathbf{n}_i \otimes \mathbf{n}_i$ and $\mathbf{P}_i^\| = \mathbf{t}_i \otimes \mathbf{t}_i$ so that the stiffness tensor takes the form $\mathbf{K}_i = K_n^i \left( \mathbf{n}_i \otimes \mathbf{n}_i \right) + K_t^i \left( \mathbf{t}_i \otimes \mathbf{t}_i \right)$ and the equation of motion (7) and (8) are specialized

$$\sum_{i=1}^{n}\left\{\left[K_n^i\left(\mathbf{n}_i\otimes\mathbf{n}_i\right)+K_t^i\left(\mathbf{t}_i\otimes\mathbf{t}_i\right)\right]\left(\mathbf{u}-\mathbf{u}_i\right)+\frac{l_i}{2}K_t^i\mathbf{t}_i\left(\phi+\phi_i\right)\right\}+\mathbf{M}\ddot{\mathbf{u}}=\mathbf{0}, \tag{9}$$

$$\sum_{i=1}^{n}\left[\frac{l_i}{2}K_t^i\mathbf{t}_i\cdot\left(\mathbf{u}-\mathbf{u}_i\right)+K_\varphi^i\left(\phi-\phi_i\right)+\frac{l_i^2}{4}K_t^i\left(\phi+\phi_i\right)\right]+J\ddot{\phi}=\mathbf{0}. \tag{10}$$

A system of three ODEs is obtained for each rigid block in terms of the generalized displacements. If the propagation of harmonic plane waves is analyzed, travelling along axis $\mathbf{i}$, the generalized displacement field of the block with centre located at $\mathbf{x}$ is assumed in the standard form $\mathbf{U}=\hat{\mathbf{U}}\exp\left[i\left(\mathbf{k}\cdot\mathbf{x}-\omega t\right)\right]$, where $\mathbf{k}=q\mathbf{i}$ is the wave vector and $q$ and $\omega$ denote the wave number and the circular frequency, respectively, and $\hat{\mathbf{U}}=\left\{\hat{\mathbf{u}}^T\ \hat{\phi}\right\}^T=\left\{\hat{u}_1\ \hat{u}_2\ \hat{\phi}\right\}^T$ is the polarization vector. Substituting the assumed generalized displacement field in the equation of motion, one obtains the algebraic system of three homogeneous linear equations having as unknown the polarization vector:

$$\sum_{i=1}^{n}\left\{\left[1-\exp\left(i\mathbf{k}\cdot\mathbf{x}_i\right)\right]\mathbf{K}_i\hat{\mathbf{u}}+\frac{l_i}{2}\left[1+\exp\left(i\mathbf{k}\cdot\mathbf{x}_i\right)\right]\mathbf{K}_i\mathbf{t}_i\hat{\phi}\right\}-\omega^2\mathbf{M}\hat{\mathbf{u}}=\mathbf{0},$$

$$\sum_{i=1}^{n}\left\{\begin{array}{l}\frac{l_i}{2}\left[1-\exp\left(i\mathbf{k}\cdot\mathbf{x}_i\right)\right]\mathbf{K}_i\mathbf{t}_i\cdot\hat{\mathbf{u}}+\\ +\left[K_\varphi^i\left[1-\exp\left(i\mathbf{k}\cdot\mathbf{x}_i\right)\right]+\frac{l_i^2}{4}\mathbf{t}_i\cdot\mathbf{K}_i\mathbf{t}_i\left[1+\exp\left(i\mathbf{k}\cdot\mathbf{x}_i\right)\right]\right]\hat{\phi}\end{array}\right\}-\omega^2J\hat{\phi}=\mathbf{0}. \tag{11}$$

Since a centro-symmetric microstructure of the blocky material is considered, to each vector $\mathbf{x}_i$ (connecting the center of the reference block to the center of the $i$-th adjacent one) corresponds an opposite vector $\mathbf{x}_j=-\mathbf{x}_i$ identifying the $j$-th block, with $\mathbf{t}_j=-\mathbf{t}_i$, so that the following property $\mathbf{K}_j=\mathbf{K}_i$ applies. Therefore, one obtains the following eigen-problem in the compact form

$$\left[\mathbf{C}_{Lag}(\mathbf{k})-\omega^2\mathbf{M}\right]\hat{\mathbf{U}}=\left\{\begin{bmatrix}\mathbf{A}&+i\mathbf{b}\\-i\mathbf{b}^T&C\end{bmatrix}-\omega^2\begin{bmatrix}M\mathbf{I}_2&\mathbf{0}\\\mathbf{0}&J\end{bmatrix}\right\}\begin{Bmatrix}\hat{\mathbf{u}}\\\hat{\phi}\end{Bmatrix}=\mathbf{0}, \tag{12}$$

with

$$\mathbf{A} = \sum_{i=1}^{n} \left[1 - \cos(\mathbf{k} \cdot \mathbf{x}_i)\right] \mathbf{K}_i \quad ,$$

$$\mathbf{b} = \sum_{i=1}^{n} \left[\frac{l_i}{2} \sin(\mathbf{k} \cdot \mathbf{x}_i) \mathbf{K}_i \mathbf{t}_i\right] \quad , \tag{13}$$

$$C = \sum_{i=1}^{n} \left\{ K_\varphi^i \left[1 - \cos(\mathbf{k} \cdot \mathbf{x}_i)\right] + \frac{l_i^2}{4} \mathbf{t}_i \cdot \mathbf{K}_i \mathbf{t}_i \left[1 + \cos(\mathbf{k} \cdot \mathbf{x}_i)\right] \right\} \quad .$$

Moreover, if the interfaces are normal to the connecting vector, namely $\mathbf{d}_i = \mathbf{n}_i$, $i = 1, n$, the above terms take a simpler form:

$$\mathbf{A} = \sum_{i=1}^{n} \left[1 - \cos(\mathbf{k} \cdot \mathbf{x}_i)\right] \left[K_n^i (\mathbf{n}_i \otimes \mathbf{n}_i) + K_t^i (\mathbf{t}_i \otimes \mathbf{t}_i)\right] \quad ,$$

$$\mathbf{b} = \sum_{i=1}^{n} \left[\frac{l_i}{2} K_t^i \sin(\mathbf{k} \cdot \mathbf{x}_i) \mathbf{t}_i\right] \quad , \tag{14}$$

$$C = \sum_{i=1}^{n} \left\{ K_\varphi^i \left[1 - \cos(\mathbf{k} \cdot \mathbf{x}_i)\right] + \frac{l_i^2}{4} K_t^i \left[1 + \cos(\mathbf{k} \cdot \mathbf{x}_i)\right] \right\} \quad .$$

For a given wave vector $\mathbf{k}$ the corresponding angular frequency $\omega(\mathbf{k})$ is obtained as solution of the eigen-problem (12) with three solutions each of them defining a dispersive branch in the Floquet-Bloch spectrum. When considering the long waves asymptotic, namely for $|\mathbf{k}| \to 0$, from (13) one obtains $\mathbf{A} = \mathbf{0}$, $\mathbf{a} = \mathbf{b} = \mathbf{0}$, while $C_0 = C(|\mathbf{k}| \to 0) = \sum_{i=1}^{n} \left(\frac{l_i^2}{2} \mathbf{t}_i \cdot \mathbf{K}_i \mathbf{t}_i\right)$. In this limit case, two vanishing frequencies are obtained $\omega_{ac1,2}(|\mathbf{k}| \to 0) = 0$, from which two acoustic branches depart, and a critical point $\omega_{opt}(|\mathbf{k}| = 0) = \sqrt{\frac{C}{J}} = \sqrt{\sum_{i=1}^{n} \left(\frac{l_i^2}{2} \mathbf{t}_i \cdot \mathbf{K}_i \mathbf{t}_i\right) / J}$ with vanishing group velocity, representing an extreme of the optical branch.

### 3. Micropolar dynamic homogenization of periodic centro-symmetric blocky systems

The equation of motion of the discrete Lagrangian model may be approximated through a continuum model in which the generalized displacements are depending on two continuum fields $\mathbf{v}(\mathbf{x})$ and $\theta(\mathbf{x})$ playing the role of generalized macro-displacements. This objective may be achieved by assuming a down-scaling law, namely an approximation of the displacement vector and the rotation of the *i*-th ring through a second order expansion of the macro-displacement and rotation fields

$$\mathbf{u}_i(t) \cong \mathbf{v}(\mathbf{x},t) + l_i \mathbf{H}(\mathbf{x},t)\mathbf{n}_i + \frac{1}{2}l_i^2 \nabla \mathbf{H}(\mathbf{x},t) : (\mathbf{n}_i \otimes \mathbf{n}_i) ,$$

$$\phi_i(t) \cong \theta(\mathbf{x},t) + l_i \boldsymbol{\chi}(\mathbf{x},t) \cdot \mathbf{n}_i + \frac{1}{2}l_i^2 \nabla \boldsymbol{\chi}(\mathbf{x},t) : (\mathbf{n}_i \otimes \mathbf{n}_i) ,$$

(15)

$\mathbf{H} = \nabla \mathbf{v}$ and $\nabla \mathbf{H}$ being the macro-displacement gradient and second gradient, $\boldsymbol{\chi} = \nabla \theta$ and $\nabla \boldsymbol{\chi}$ the curvature and its gradient tensor, respectively.

*3.1 Continualization of the discrete equations of motion*

The block generalized displacements in the equation of motion (7) and (8) of the reference block are now substituted by the second order expansion (15) so obtaining three PDEs

$$\sum_{i=1}^{n} \left[ \mathbf{K}_i \left( l_i \mathbf{H}\mathbf{n}_i + \frac{1}{2}l_i^2 \nabla \mathbf{H} : (\mathbf{n}_i \otimes \mathbf{n}_i) \right) - \frac{l_i}{2} \mathbf{K}_i \mathbf{t}_i \left( 2\theta + l_i \boldsymbol{\chi} \cdot \mathbf{n}_i + \frac{1}{2}l_i^2 \nabla \boldsymbol{\chi} : (\mathbf{n}_i \otimes \mathbf{n}_i) \right) \right] - M\ddot{\mathbf{v}} = \mathbf{0} ,$$

$$\sum_{i=1}^{n} \left\{ \begin{array}{l} \frac{l_i}{2} \mathbf{t}_i \cdot \mathbf{K}_i \left( l_i \mathbf{H}\mathbf{n}_i + \frac{1}{2}l_i^2 \nabla \mathbf{H} : (\mathbf{n}_i \otimes \mathbf{n}_i) \right) + K_\varphi^i \left[ l_i \boldsymbol{\chi} \cdot \mathbf{n}_i + \frac{1}{2}l_i^2 \nabla \boldsymbol{\chi} : (\mathbf{n}_i \otimes \mathbf{n}_i) \right] + \\ -\frac{l_i^2}{4} \mathbf{t}_i \cdot \mathbf{K}_i \mathbf{t}_i \left[ 2\theta + l_i \boldsymbol{\chi} \cdot \mathbf{n}_i + \frac{1}{2}l_i^2 \nabla \boldsymbol{\chi} : (\mathbf{n}_i \otimes \mathbf{n}_i) \right] \end{array} \right\} - J\ddot{\theta} = 0 ,$$

(16)

in terms of $\mathbf{v}(\mathbf{x})$ and $\theta(\mathbf{x})$ and of their gradients. It is worth to note that because the centro-symmetry of the blocks, some terms in in equation (16) are vanishing being sums of odd functions of the unit vectors $\mathbf{n}_i$ and $\mathbf{t}_i$, i.e. $\sum_{i=1}^{n} \left[ l_i \mathbf{K}_i \mathbf{H} \mathbf{n}_i \right] = \mathbf{0}$, $\sum_{i=1}^{n} \left[ l_i \mathbf{K}_i \mathbf{t}_i \theta \right] = \mathbf{0}$,

$\sum_{i=1}^{n} \left\{ \frac{l_i^3}{4} \left[ \nabla \boldsymbol{\chi} : (\mathbf{n}_i \otimes \mathbf{n}_i) \right] \mathbf{K}_i \mathbf{t}_i \right\} = \mathbf{0}$, $\sum_{i=1}^{n} \left\{ \frac{l_i^3}{4} (\mathbf{t}_i \cdot \mathbf{K}_i) \left[ \nabla \mathbf{H} : (\mathbf{n}_i \otimes \mathbf{n}_i) \right] \right\} = \mathbf{0}$, $\sum_{i=1}^{n} \left\{ K_\varphi^i l_i \boldsymbol{\chi} \cdot \mathbf{n}_i \right\} = 0$,

$\sum_{i=1}^{n} \left\{ \frac{l_i^3}{4} (\mathbf{t}_i \cdot \mathbf{K}_i \mathbf{t}_i) (\boldsymbol{\chi} \cdot \mathbf{n}_i) \right\} = 0$. Therefore, equation (16) takes the following simpler form

$$\sum_{i=1}^{n}\left[\frac{1}{2}l_i^2\mathbf{K}_i\left(\nabla\mathbf{H}:(\mathbf{n}_i\otimes\mathbf{n}_i)\right)-\frac{l_i^2}{2}(\boldsymbol{\chi}\cdot\mathbf{n}_i)\mathbf{K}_i\mathbf{t}_i\right]-M\ddot{\mathbf{v}}=\mathbf{0} \quad,$$

$$\sum_{i=1}^{n}\left\{\begin{array}{l}\frac{l_i^2}{2}\mathbf{t}_i\cdot\mathbf{K}_i(\mathbf{H}\mathbf{n}_i)+\frac{l_i^2}{2}K_\varphi^i\left[\nabla\boldsymbol{\chi}:(\mathbf{n}_i\otimes\mathbf{n}_i)\right]+\\ -\frac{l_i^2}{2}\mathbf{t}_i\cdot\mathbf{K}_i\mathbf{t}_i\theta-\frac{l_i^4}{8}(\mathbf{t}_i\cdot\mathbf{K}_i\mathbf{t}_i)\left[\nabla\boldsymbol{\chi}:(\mathbf{n}_i\otimes\mathbf{n}_i)\right]\end{array}\right\}-J\ddot{\theta}=0 \quad,$$

(17)

and, after a rearrangement of the terms

$$\sum_{i=1}^{n}\frac{l_i^2}{2}\left[(\mathbf{K}_i\otimes\mathbf{n}_i\otimes\mathbf{n}_i)\vdots\nabla\mathbf{H}-(\mathbf{K}_i\mathbf{t}_i\otimes\mathbf{n}_i)\boldsymbol{\chi}\right]-M\ddot{\mathbf{v}}=\mathbf{0} \quad,$$

$$\sum_{i=1}^{n}\frac{l_i^2}{2}\left\{(\mathbf{K}_i\mathbf{t}_i\otimes\mathbf{n}_i):\mathbf{H}-\mathbf{t}_i\cdot\mathbf{K}_i\mathbf{t}_i\theta+\left[K_\varphi^i-\frac{l_i^2}{4}(\mathbf{t}_i\cdot\mathbf{K}_i\mathbf{t}_i)\right](\mathbf{n}_i\otimes\mathbf{n}_i):\nabla\boldsymbol{\chi}\right\}-J\ddot{\theta}=0 \quad.$$

(18)

From equation (5) the following relation holds

$$(\mathbf{K}_i\otimes\mathbf{n}_i\otimes\mathbf{n}_i)\vdots\nabla\mathbf{H}=$$
$$=\begin{bmatrix}\left(K_n^i\cos^2\beta_i+K_t^i\sin^2\beta_i\right)\mathbf{n}_i\otimes\mathbf{n}_i\otimes\mathbf{n}_i\otimes\mathbf{n}_i+\\ +\left(K_n^i-K_t^i\right)\sin 2\beta_i \ \mathrm{sym}(\mathbf{t}_i\otimes\mathbf{n}_i)\otimes\mathbf{n}_i\otimes\mathbf{n}_i+\\ +\left(K_n^i\sin^2\beta_i+K_t^i\cos^2\beta_i\right)\mathbf{t}_i\otimes\mathbf{t}_i\otimes\mathbf{n}_i\otimes\mathbf{n}_i\end{bmatrix}\vdots\nabla\mathbf{H} \quad.$$

(19)

On the other side, the term in equation (18) may be written as

$$(\mathbf{K}_i\mathbf{t}_i\otimes\mathbf{n}_i)\boldsymbol{\chi}=\begin{bmatrix}\left(K_n^i-K_t^i\right)\sin\beta_i\cos\beta_i \ \mathbf{t}_i\otimes\mathbf{n}_i\otimes\mathbf{n}_i\otimes\mathbf{n}_i+\\ +\left(K_n^i\sin^2\beta_i+K_t^i\cos^2\beta_i\right)\mathbf{t}_i\otimes\mathbf{t}_i\otimes\mathbf{n}_i\otimes\mathbf{n}_i\end{bmatrix}\vdots\nabla\mathbf{W} \quad,$$

(20)

where the asymmetric tensor of rotation $\mathbf{W}$ is introduced, with components $w_{jh}=-\in_{3jh}\theta$, $\in_{jkl}$ being the Levi-Civita symbol, having the following properties $\theta=\mathbf{t}_i\cdot\mathbf{W}\mathbf{n}_i$, $\mathbf{n}_i\cdot\mathbf{W}\mathbf{n}_i=0$, $(\mathbf{n}_i\otimes\mathbf{n}_i\otimes\mathbf{n}_i\otimes\mathbf{n}_i)\vdots\nabla\mathbf{W}=\mathbf{0}$ and $(\mathbf{t}_i\otimes\mathbf{n}_i\otimes\mathbf{n}_i\otimes\mathbf{n}_i)\vdots\nabla\mathbf{W}=\mathbf{0}$. By comparing equation (20) with (19) it follows that

$$(\mathbf{K}_i\mathbf{t}_i\otimes\mathbf{n}_i)\boldsymbol{\chi}=\begin{bmatrix}\left(K_n^i\cos^2\beta_i+K_t^i\sin^2\beta_i\right)\mathbf{n}_i\otimes\mathbf{n}_i\otimes\mathbf{n}_i\otimes\mathbf{n}_i+\\ +\left(K_n^i-K_t^i\right)\sin 2\beta_i \ \mathrm{sym}(\mathbf{t}_i\otimes\mathbf{n}_i)\otimes\mathbf{n}_i\otimes\mathbf{n}_i+\\ +\left(K_n^i\sin^2\beta_i+K_t^i\cos^2\beta_i\right)\mathbf{t}_i\otimes\mathbf{t}_i\otimes\mathbf{n}_i\otimes\mathbf{n}_i\end{bmatrix}\vdots\nabla\mathbf{W} \quad.$$

(21)

If the asymmetric strain tensor of Cosserat $\mathbf{\Gamma} = \mathbf{H} - \mathbf{W}(\theta)$ is now introduced, the first equation in (18) may be rewritten in the form

$$\sum_{i=1}^{n} \frac{l_i^2}{2} \begin{bmatrix} \left(K_n^i \cos^2 \beta_i + K_t^i \sin^2 \beta_i\right) \mathbf{n}_i \otimes \mathbf{n}_i \otimes \mathbf{n}_i \otimes \mathbf{n}_i + \\ +\left(K_n^i - K_t^i\right) \sin 2\beta_i \ \mathrm{sym}\left(\mathbf{t}_i \otimes \mathbf{n}_i\right) \otimes \mathbf{n}_i \otimes \mathbf{n}_i + \\ +\left(K_n^i \sin^2 \beta_i + K_t^i \cos^2 \beta_i\right) \mathbf{t}_i \otimes \mathbf{t}_i \otimes \mathbf{n}_i \otimes \mathbf{n}_i \end{bmatrix} : \nabla \mathbf{\Gamma} - \mathbf{M} \ddot{\mathbf{v}} = \mathbf{0} \ . \quad (22)$$

Moreover, once noted that the following relations apply $\left(\mathbf{t}_i \otimes \mathbf{t}_i \otimes \mathbf{n}_i \otimes \mathbf{n}_i\right) : \nabla \mathbf{\Gamma} = \nabla \cdot \left[\left(\mathbf{t}_i \otimes \mathbf{n}_i \otimes \mathbf{t}_i \otimes \mathbf{n}_i\right) : \mathbf{\Gamma}\right]$ and $\left[\mathrm{sym}\left(\mathbf{t}_i \otimes \mathbf{n}_i\right) \otimes \mathbf{n}_i \otimes \mathbf{n}_i\right] : \nabla \mathbf{\Gamma} = \nabla \cdot \left[\left(\mathbf{t}_i \otimes \mathbf{n}_i \otimes \mathbf{n}_i \otimes \mathbf{n}_i + \mathbf{n}_i \otimes \mathbf{n}_i \otimes \mathbf{t}_i \otimes \mathbf{n}_i\right) : \mathbf{\Gamma}\right]$, the equation of motion (22) may be rewritten in the compact form

$$\nabla \cdot (\mathbb{E}_s \mathbf{\Gamma}) = \rho \ddot{\mathbf{v}} \ , \quad (23)$$

having defined the fourth order elastic tensor

$$\mathbb{E}_s = \frac{1}{A_b} \sum_{i=1}^{n} \frac{l_i^2}{2} \begin{bmatrix} \left(K_n^i \cos^2 \beta_i + K_t^i \sin^2 \beta_i\right) \mathbf{n}_i \otimes \mathbf{n}_i \otimes \mathbf{n}_i \otimes \mathbf{n}_i + \\ +\frac{\left(K_n^i - K_t^i\right)}{2} \sin 2\beta_i \ \left(\mathbf{t}_i \otimes \mathbf{n}_i \otimes \mathbf{n}_i \otimes \mathbf{n}_i + \mathbf{n}_i \otimes \mathbf{n}_i \otimes \mathbf{t}_i \otimes \mathbf{n}_i\right) + \\ +\left(K_n^i \sin^2 \beta_i + K_t^i \cos^2 \beta_i\right) \mathbf{t}_i \otimes \mathbf{n}_i \otimes \mathbf{t}_i \otimes \mathbf{n}_i \end{bmatrix} , \quad (24)$$

provided with major symmetry, being $A_b$ the area and $\rho = M/A_b$ the mass density of the current block, respectively. From equation (23) the asymmetric stress tensor

$$\mathbf{T} = \mathbb{E}_s \mathbf{\Gamma} , \quad (25)$$

of the equivalent homogeneous continuum may be identified, with components

$$\sigma_{hk} = \frac{1}{A_b} \sum_{i=1}^{n} \frac{l_i^2}{2} \begin{bmatrix} \left(K_n^i \cos^2 \beta_i + K_t^i \sin^2 \beta_i\right) \left(n_h^i n_k^i n_p^i n_q^i\right) + \\ +\frac{\left(K_n^i - K_t^i\right)}{2} \sin 2\beta_i \ \left(t_h^i n_k^i n_p^i n_q^i + n_h^i n_k^i t_p^i n_q^i\right) + \\ +\left(K_n^i \sin^2 \beta_i + K_t^i \cos^2 \beta_i\right) \left(t_h^i n_k^i t_p^i n_q^i\right) \end{bmatrix} \gamma_{pq} \ . \quad (26)$$

Let us consider now the difference of the asymmetric stress components

$$\sigma_{21} - \sigma_{12} = - \in_{3jh} \left(\mathbf{e}_j \otimes \mathbf{e}_h\right) : \left(\mathbb{E}_s \mathbf{\Gamma}\right) =$$
$$= \frac{1}{A_b} \sum_{i=1}^{n} \frac{l_i^2}{2} \left[ \left(K_n^i - K_t^i\right) \sin \beta_i \cos \beta_i \left(\mathbf{n}_i \otimes \mathbf{n}_i\right) + \left(K_n^i \sin^2 \beta_i + K_t^i \cos^2 \beta_i\right) \left(\mathbf{t}_i \otimes \mathbf{n}_i\right)\right] = \quad (27)$$
$$= \frac{1}{A_b} \sum_{i=1}^{n} \frac{l_i^2}{2} \left[\left(\mathbf{K}_i \mathbf{t}_i \otimes \mathbf{n}_i\right) : \mathbf{H} - \mathbf{t}_i \cdot \mathbf{K}_i \mathbf{t}_i \theta\right]$$

and define the symmetric second order elastic tensor vector

$$\mathbf{E}_s = \frac{1}{A_b} \sum_{i=1}^{n} \frac{l_i^2}{2} \left[ K_\varphi^i - \frac{l_i^2}{4} \left(\mathbf{t}_i \cdot \mathbf{K}_i \mathbf{t}_i\right)\right] \left(\mathbf{n}_i \otimes \mathbf{n}_i\right) =$$
$$= \frac{1}{A_b} \sum_{i=1}^{n} \frac{l_i^2}{2} \left[ K_\varphi^i - \frac{l_i^2}{4} \left(K_n^i \sin^2 \beta_i + K_t^i \cos^2 \beta_i\right)\right] \left(\mathbf{n}_i \otimes \mathbf{n}_i\right) \quad (28)$$

By comparing equations (27) and (28) with second equation of motion (18), this one takes the following compact form

$$\nabla \cdot \left(\mathbf{E}_s \, \mathbf{\chi}\right) - \in_{3jh} \left(\mathbf{e}_j \otimes \mathbf{e}_h\right) : \left(\mathbb{E}_s \mathbf{\Gamma}\right) = I \ddot{\theta} \,, \quad j, h = 1, 2 \,, \quad (29)$$

where the density of rotational inertia is defined $I = J/A_{cell} = \rho r^2$. If the vector of couple-stress

$$\mathbf{m} = \mathbf{E}_s \, \mathbf{\chi} \,, \quad (30)$$

is recognized, the equation of motion of the homogeneous micropolar continuum equivalent to the blocky material is obtained in terms of the asymmetric stress tensor and of the couple-stress vector as follows

$$\nabla \cdot \mathbf{T} = \rho \ddot{\mathbf{v}} \,,$$
$$\nabla \cdot \mathbf{m} - \in_{3jh} \left(\mathbf{e}_j \otimes \mathbf{e}_h\right) : \mathbf{T} = I \ddot{\theta} \,. \quad (31)$$

The resulting equivalent micropolar continuum is represented by the generalized displacement $\mathbf{v}(\mathbf{x})$ and $\theta(\mathbf{x})$, the asymmetric strain tensor $\mathbf{\Gamma}(\mathbf{x})$ and the in-plane curvature vector $\mathbf{\chi}(\mathbf{x})$, the asymmetric stress tensor $\mathbf{T}(\mathbf{x})$ and the couple-stress vector $\mathbf{m}(\mathbf{x})$ with equation of motion given in the form (23) and (29). Because the centro-symmetry of the blocks, the strain tensor and the curvature vector are uncoupled. Moreover, the positive definiteness of the second order tensor $\mathbf{E}_s$ is not guarantee, as may be observed by definition (28). This outcome may be observed also in Sulem and Mühlhaus, 1997, equation (10), where the stiffnesses associated to the curvatures are negative because the rotational stiffness term of the interfaces $K_\varphi^i$ was assumed vanishing.

*3.2 Variational derivation of the micropolar model*

The results obtained in the previous Section, here are achieved by following a different approach based on an extended Hamiltonian derivation of the equations of motion for the homogeneous continuum equivalent to the blocky material. To this end, let consider a cluster $C$ of periodic blocks having characteristic size $L \gg l$ with respect to the characteristic block size $l$. Since dissipative mechanisms are ignored, the Lagrangian function $\mathcal{L}_C = T_C - \Pi_C$ associated to the blocky cluster is considered, with the kinetic energy written in the form

$$T_C = \int_C \left[ \frac{1}{2}\rho \dot{\mathbf{v}} \cdot \dot{\mathbf{v}} + \frac{1}{2} I \dot{\theta}^2 \right] da \quad , \tag{32}$$

being the average mass density $\rho$ and the density of rotational inertia $I$ defined in the previous Section. The strain energy $\Pi_C$ stored in the cluster is obtained as the superposition of the elastic potential energy stored in the interfaces.

The elastic potential energy stored in the interface connecting the reference block to the $i$-th adjacent one centered at $\mathbf{x}_i$ is approximated through a the second order expansion of the generalized displacement field (15). Accordingly, the relative generalized displacement between the adjacent block at the interface is approximated as follows

$$\Delta \mathbf{s}_i = (\mathbf{u}_i - \mathbf{u}) - (\boldsymbol{\phi} + \boldsymbol{\phi}_i)\frac{l_i}{2}\mathbf{t}_i =$$
$$= l_i \mathbf{H} \mathbf{n}_i + \frac{1}{2}l_i^2 \nabla \mathbf{H} : (\mathbf{n}_i \otimes \mathbf{n}_i) - \left[ 2\theta + l_i \boldsymbol{\chi} \cdot \mathbf{n}_i + \frac{1}{2}l_i^2 \nabla \boldsymbol{\chi} : (\mathbf{n}_i \otimes \mathbf{n}_i) \right]\frac{l_i}{2}\mathbf{t}_i \quad , \tag{33}$$
$$\Delta \varphi_i = \phi_i - \phi = l_i \boldsymbol{\chi}(\mathbf{x},t) \cdot \mathbf{n}_i + \frac{1}{2}l_i^2 \nabla \boldsymbol{\chi}(\mathbf{x},t) : (\mathbf{n}_i \otimes \mathbf{n}_i) \quad .$$

After noting that $\mathbf{H}\mathbf{n}_i - \theta \mathbf{t}_i = \left[ (\mathbf{n}_i \otimes \mathbf{n}_i) : \boldsymbol{\Gamma} \right] \mathbf{n}_i + \left[ (\mathbf{t}_i \otimes \mathbf{n}_i) : \boldsymbol{\Gamma} \right] \mathbf{t}_i$, the relative displacement may be written as

$$\Delta \mathbf{s}_i = l_i \left[ (\mathbf{n}_i \otimes \mathbf{n}_i) : \mathbf{\Gamma} \right] \mathbf{n}_i + l_i \left[ (\mathbf{t}_i \otimes \mathbf{n}_i) : \mathbf{\Gamma} \right] \mathbf{t}_i +$$
$$+ \frac{1}{2} l_i^2 (\mathbf{n}_i \otimes \mathbf{n}_i) : \nabla \mathbf{H} - \frac{l_i^2}{2} (\boldsymbol{\chi} \cdot \mathbf{n}_i) \mathbf{t}_i - \frac{l_i^3}{4} \left[ \nabla \boldsymbol{\chi} : (\mathbf{n}_i \otimes \mathbf{n}_i) \right] \mathbf{t}_i \quad . \tag{34}$$

The total elastic potential energy stored in the cluster is obtained from equation (1) by noting that summing up all the contributions, for each couple of adjacent blocks, the energy is doubled, so that one obtains

$$\Pi_C = \frac{1}{2} \int_C \frac{1}{2 A_b} \sum_{i=1}^{n} \left[ \Delta \mathbf{s}_i \cdot \mathbf{K}_i \Delta \mathbf{s}_i + K_\varphi^i \Delta \varphi_i^2 \right] da \quad . \tag{35}$$

When substituting (34) and the second of (33) in (35), some terms result in a dyadic product of an odd number of vectors $\mathbf{n}_i$ and $\mathbf{t}_i$, whose sums over $n$ are zero because the assumption od centro-symmetric blocks. Moreover, since the present analysis limited to the formulation of a classical micropolar model (see Eremeyev *et al.*, 2013), namely with the elastic potential energy only depending on the first gradients of the generalized macro-displacements field, the terms in (35) depending on the second gradient terms $\nabla \mathbf{H}$ and $\nabla \boldsymbol{\chi}$ are disregarded, apart from the terms $\left[ (\mathbf{t}_i \otimes \mathbf{n}_i) : \mathbf{W} \right] \left[ \nabla \boldsymbol{\chi} : (\mathbf{n}_i \otimes \mathbf{n}_i) \right]$ involving the product of $\mathbf{W}$ ($\mathbf{\Gamma} = \mathbf{H} - \mathbf{W}$) and $\nabla \boldsymbol{\chi}$. Therefore, the total elastic potential energy takes the form

$$\Pi_C = \frac{1}{4 A_b} \int_C \sum_{i=1}^{n} \left\{ \begin{array}{l} l_i^2 \left[ (\mathbf{n}_i \otimes \mathbf{n}_i) : \mathbf{\Gamma} \right]^2 (\mathbf{n}_i \cdot \mathbf{K}_i \mathbf{n}_i) + l_i^2 \left[ (\mathbf{t}_i \otimes \mathbf{n}_i) : \mathbf{\Gamma} \right]^2 (\mathbf{t}_i \cdot \mathbf{K}_i \mathbf{t}_i) + \\ + 2 l_i^2 \left[ (\mathbf{n}_i \otimes \mathbf{n}_i) : \mathbf{\Gamma} \right] \left[ (\mathbf{t}_i \otimes \mathbf{n}_i) : \mathbf{\Gamma} \right] (\mathbf{n}_i \cdot \mathbf{K}_i \mathbf{t}_i) + \\ + \frac{l_i^4}{4} (\boldsymbol{\chi} \cdot \mathbf{n}_i)^2 (\mathbf{t}_i \cdot \mathbf{K}_i \mathbf{t}_i) + K_\varphi^i l_i^2 (\boldsymbol{\chi} \cdot \mathbf{n}_i)^2 + \\ + \frac{l_i^4}{2} \left[ (\mathbf{n}_i \otimes \mathbf{n}_i) : \mathbf{W} \right] \left[ \nabla \boldsymbol{\chi} : (\mathbf{n}_i \otimes \mathbf{n}_i) \right] (\mathbf{n}_i \cdot \mathbf{K}_i \mathbf{t}_i) + \\ + \frac{l_i^4}{2} \left[ (\mathbf{t}_i \otimes \mathbf{n}_i) : \mathbf{W} \right] \left[ \nabla \boldsymbol{\chi} : (\mathbf{n}_i \otimes \mathbf{n}_i) \right] (\mathbf{t}_i \cdot \mathbf{K}_i \mathbf{t}_i) \end{array} \right\} da \quad . \tag{36}$$

Let us consider now the first five terms of the summation over the coordination number $n$ in (36) and note that

$$l_i^2 \left[ (\mathbf{n}_i \otimes \mathbf{n}_i) : \boldsymbol{\Gamma} \right]^2 (\mathbf{n}_i \cdot \mathbf{K}_i \mathbf{n}_i) + l_i^2 \left[ (\mathbf{t}_i \otimes \mathbf{n}_i) : \boldsymbol{\Gamma} \right]^2 (\mathbf{t}_i \cdot \mathbf{K}_i \mathbf{t}_i) +$$

$$+ 2 l_i^2 \left[ (\mathbf{n}_i \otimes \mathbf{n}_i) : \boldsymbol{\Gamma} \right] \left[ (\mathbf{t}_i \otimes \mathbf{n}_i) : \boldsymbol{\Gamma} \right] (\mathbf{n}_i \cdot \mathbf{K}_i \mathbf{t}_i) + l_i^2 \left[ K_\varphi^i + \frac{l_i^2}{4} (\mathbf{t}_i \cdot \mathbf{K}_i \mathbf{t}_i) \right] (\boldsymbol{\chi} \bullet \mathbf{n}_i)^2 =$$

$$= \boldsymbol{\Gamma} : \left\{ \begin{array}{l} \left( K_n^i \cos^2 \beta_i + K_t^i \sin^2 \beta_i \right) l_i^2 (\mathbf{n}_i \otimes \mathbf{n}_i \otimes \mathbf{n}_i \otimes \mathbf{n}_i) + \\ + \left( K_n^i \sin^2 \beta_i + K_t^i \cos^2 \beta_i \right) l_i^2 (\mathbf{t}_i \otimes \mathbf{n}_i \otimes \mathbf{t}_i \otimes \mathbf{n}_i) + \\ + \frac{(K_n^i - K_t^i)}{2} \sin 2\beta_i l_i^2 \left[ (\mathbf{t}_i \otimes \mathbf{n}_i \otimes \mathbf{n}_i \otimes \mathbf{n}_i) + (\mathbf{n}_i \otimes \mathbf{n}_i \otimes \mathbf{t}_i \otimes \mathbf{n}_i) \right] \end{array} \right\} \boldsymbol{\Gamma} +$$

$$+ l_i^2 \left[ K_\varphi^i + \frac{l_i^2}{4} (\mathbf{t}_i \cdot \mathbf{K}_i \mathbf{t}_i) \right] \boldsymbol{\chi} \bullet (\mathbf{n}_i \otimes \mathbf{n}_i) \boldsymbol{\chi} \quad ,$$

while the sixth term is zero because $(\mathbf{n}_i \otimes \mathbf{n}_i) : \mathbf{W} = 0$. Being $(\mathbf{t}_i \otimes \mathbf{n}_i) : \mathbf{W} = \theta$, in the seventh term one notes that $\left[ (\mathbf{t}_i \otimes \mathbf{n}_i) : \mathbf{W} \right] \left[ \nabla \boldsymbol{\chi} : (\mathbf{n}_i \otimes \mathbf{n}_i) \right] = \theta \, \nabla \boldsymbol{\chi} : (\mathbf{n}_i \otimes \mathbf{n}_i) = \theta \, \theta_{,pq} n_p^i n_q^i$. By an application of the Divergence theorem (see Kumar and McDowell (2004)) the seventh term becomes

$$\sum_i^N l_i^4 (\mathbf{t}_i \cdot \mathbf{K}_i \mathbf{t}_i) n_p^i n_q^i \int_C \theta \, \theta_{,pq} \, da = -\sum_{i=1}^N l_i^4 (\mathbf{t}_i \cdot \mathbf{K}_i \mathbf{t}_i) n_p^i n_q^i \int_C \theta_{,p} \, \theta_{,q} \, da +$$

$$+ \sum_{i=1}^N l_i^4 (\mathbf{t}_i \cdot \mathbf{K}_i \mathbf{t}_i) n_p^i n_q^i \int_{\partial C} \theta_{,p} \, \theta \, v_q \, ds = -\sum_{i=1}^N \int_C l_i^4 (\mathbf{t}_i \cdot \mathbf{K}_i \mathbf{t}_i) (\boldsymbol{\chi} \cdot \mathbf{n}_i)^2 \, da + \text{boundary terms} \quad ,$$

where the boundary terms are defined on the boundary of $C$. The total elastic potential energy in the cluster of blocks is written in the following form

$$\Pi_C = \frac{1}{2 A_b} \int_C \left\{ \boldsymbol{\Gamma} : \sum_{i=1}^n \frac{l_i^2}{2 A_{cell}} \left[ \begin{array}{l} \left( K_n^i \cos^2 \beta_i + K_t^i \sin^2 \beta_i \right) (\mathbf{n}_i \otimes \mathbf{n}_i \otimes \mathbf{n}_i \otimes \mathbf{n}_i) + \\ + \frac{(K_n^i - K_t^i)}{2} \sin 2\beta_i \left[ (\mathbf{t}_i \otimes \mathbf{n}_i \otimes \mathbf{n}_i \otimes \mathbf{n}_i) + (\mathbf{n}_i \otimes \mathbf{n}_i \otimes \mathbf{t}_i \otimes \mathbf{n}_i) \right] \\ + \left( K_n^i \sin^2 \beta_i + K_t^i \cos^2 \beta_i \right) (\mathbf{t}_i \otimes \mathbf{n}_i \otimes \mathbf{t}_i \otimes \mathbf{n}_i) + \end{array} \right] \boldsymbol{\Gamma} + \right.$$

$$\left. + \boldsymbol{\chi} \bullet \sum_{i=1}^n \frac{l_i^2}{2 A_{cell}} \left[ K_\varphi^i - \frac{l_i^2}{4} \left( K_n^i \sin^2 \beta_i + K_t^i \cos^2 \beta_i \right) \right] (\mathbf{n}_i \otimes \mathbf{n}_i) \boldsymbol{\chi} \right\} da + \quad (37)$$

that results in the classical form of the centrosymmetric micropolar continuum

$$\Pi_C = \frac{1}{2} \int_C (\boldsymbol{\Gamma} : \mathbb{E}_s \boldsymbol{\Gamma} + \boldsymbol{\chi} \cdot \mathbf{E}_s \boldsymbol{\chi}) \, da \quad , \quad (38)$$

with the elasticity tensors given by (24) and (28).

It must be remarked that assuming a first order expansion of the generalized displacements in (15) implies the same fourth order tensors given in (24), while the second order tensor, associated to the curvature vector, differs from (28) and results to be positive definite in the form

$$\mathbf{E}_s^+ = \frac{1}{A_b} \sum_{i=1}^{n} \frac{l_i^2}{2} \left[ K_\varphi^i + \frac{l_i^2}{4} (\mathbf{t}_i \cdot \mathbf{K}_i \mathbf{t}_i) \right] (\mathbf{n}_i \otimes \mathbf{n}_i) = $$
$$= \frac{1}{A_b} \sum_{i=1}^{n} \frac{l_i^2}{2} \left[ K_\varphi^i + \frac{l_i^2}{4} (K_n^i \sin^2 \beta_i + K_t^i \cos^2 \beta_i) \right] (\mathbf{n}_i \otimes \mathbf{n}_i)$$
(39)

The Lagrangian is written in the form

$$\mathcal{L} = \int_C \left[ \frac{1}{2} \rho \dot{\mathbf{v}} \cdot \dot{\mathbf{v}} + \frac{1}{2} I \dot{\theta}^2 - \frac{1}{2} \mathbf{\Gamma} \cdot \mathbb{E}_s \mathbf{\Gamma} - \frac{1}{2} \mathbf{\chi} \cdot \mathbf{E}_s \mathbf{\chi} \right] da + boundary\ terms\ ,$$
(40)

being $C$ the cluster of blocks. By applying the extended Hamilton principle, the equation of motion of the micropolar continuum are obtained independently on the boundary terms in equation (40) and are those of the micropolar continuum already obtained in equations (31).

The density $\pi$ of the elastic potential energy is derived by (38), and the constitutive equations are written

$$\mathbf{T} = \frac{\partial \pi}{\partial \mathbf{\Gamma}} = \mathbb{E}_s \mathbf{\Gamma} \quad , \quad \mathbf{m} = \frac{\partial \pi}{\partial \mathbf{\chi}} = \mathbf{E}_s \mathbf{\chi} \ ,$$
(41)

being the asymmetric stress tensor $\mathbf{T}$ with components $\sigma_{11}, \sigma_{12}, \sigma_{21}, \sigma_{22}$, the vector of couple stress $\mathbf{m}$ with components $m_1$ and $m_2$, which are energetically conjugated, respectively, to the asymmetric strain tensor $\mathbf{\Gamma}$ with components $\gamma_{11} = u_{1,1}$, $\gamma_{22} = u_{2,2}$, $\gamma_{12} = u_{1,2} + \phi$, $\gamma_{21} = u_{2,1} - \phi$ and to the curvature $\mathbf{\chi}$ with components $\chi_1 = \phi_{,1}$ and $\chi_2 = \phi_{,2}$.

## 4. Wave propagation in the micropolar blocky material

Let consider now harmonic propagation of elastic waves in the equivalent micropolar blocky material. Consequently, the generalized macro-displacement field is assumed in the form $\mathbf{V} = \hat{\mathbf{V}} \exp[i(\mathbf{k} \cdot \mathbf{x} - \omega t)]$, with $\hat{\mathbf{V}} = \{\hat{\mathbf{v}}^T \ \hat{\theta}\}$, namely $\mathbf{v} = \hat{\mathbf{v}} \exp[i(\mathbf{k} \cdot \mathbf{x} - \omega t)]$ and $\theta = \hat{\theta} \exp[i(\mathbf{k} \cdot \mathbf{x} - \omega t)]$. The equation of motion (18), remembering (19) and (20) and noting that $\mathbf{H} = \hat{\mathbf{v}} \otimes \mathbf{k} \exp[i(\mathbf{k} \cdot \mathbf{x} - \omega t)]$, $\nabla \mathbf{H} = -\hat{\mathbf{v}} \otimes \mathbf{k} \otimes \mathbf{k} \exp[i(\mathbf{k} \cdot \mathbf{x} - \omega t)]$, $\boldsymbol{\chi} = i\hat{\theta} \mathbf{k} \exp[i(\mathbf{k} \cdot \mathbf{x} - \omega t)]$, $\nabla \boldsymbol{\chi} = i\hat{\theta} \mathbf{k} \otimes \mathbf{k} \exp[i(\mathbf{k} \cdot \mathbf{x} - \omega t)]$, takes the following algebraic homogeneous form

$$\sum_{i=1}^{n} \frac{l_i^2}{2A_b} \left[ (\mathbf{n}_i \otimes \mathbf{n}_i) : (\mathbf{k} \otimes \mathbf{k}) \right] \mathbf{K}_i \hat{\mathbf{v}} + i \sum_{i=1}^{n} \frac{l_i^2}{2A_b} \left[ (\mathbf{k} \cdot \mathbf{n}_i) \mathbf{K}_i \mathbf{t}_i \right] \hat{\theta} - \omega^2 \rho \hat{\mathbf{v}} = \mathbf{0} , \qquad (42)$$

$$-i \sum_{i=1}^{n} \frac{l_i^2}{2A_b} \left[ (\mathbf{k} \cdot \mathbf{n}_i) \mathbf{K}_i \mathbf{t}_i \right] \cdot \hat{\mathbf{v}} +$$

$$+ \sum_{i=1}^{n} \frac{l_i^2}{2A_b} \left\{ \mathbf{t}_i \cdot \mathbf{K}_i \mathbf{t}_i + \left[ K_\varphi^i - \frac{l_i^2}{4} (\mathbf{t}_i \cdot \mathbf{K}_i \mathbf{t}_i) \right] \left[ (\mathbf{n}_i \otimes \mathbf{n}_i) : (\mathbf{k} \otimes \mathbf{k}) \right] \right\} \hat{\theta} - \omega^2 I \hat{\theta} = 0 , \qquad (43)$$

namely an eigenproblem $\left[ \mathbf{C}_{\text{hom}}(\mathbf{k}) - \omega^2 \mathbf{M} \right] \hat{\mathbf{U}} = \mathbf{0}$ having the same structure of problem (12), with hermitian matrix $\mathbf{C}_{\text{hom}}(\mathbf{k})$ with submatrices:

$$\mathbf{A}_{\text{hom}} = \sum_{i=1}^{n} \frac{l_i^2}{2A_b} \left[ (\mathbf{n}_i \otimes \mathbf{n}_i) : (\mathbf{k} \otimes \mathbf{k}) \right] \mathbf{K}_i ,$$

$$\mathbf{b}_{\text{hom}} = \sum_{i=1}^{n} \frac{l_i^2}{2A_b} \left[ (\mathbf{k} \cdot \mathbf{n}_i) \mathbf{K}_i \mathbf{t}_i \right] , \qquad (44)$$

$$C_{\text{hom}} = \sum_{i=1}^{n} \frac{l_i^2}{2A_b} \left\{ \mathbf{t}_i \cdot \mathbf{K}_i \mathbf{t}_i + \left[ K_\varphi^i - \frac{l_i^2}{4} (\mathbf{t}_i \cdot \mathbf{K}_i \mathbf{t}_i) \right] \left[ (\mathbf{n}_i \otimes \mathbf{n}_i) : (\mathbf{k} \otimes \mathbf{k}) \right] \right\} .$$

Because the hermitian structure of matrix $\mathbf{C}_{\text{hom}}(\mathbf{k})$, three real eigenvalues $\omega^2$ are obtained and then three dispersion functions $\omega_h(\mathbf{k})$, $h=1,3$, representative, in case of their positivity, of the harmonic wave propagation of the corresponding vector of polarization. In the long-wave asymptotic, namely for $|\mathbf{k}| \to 0$, two acoustic branches are obtained (for which $\omega_{ac1,2}(|\mathbf{k}| \to 0) = 0$), together with an optical branch departing from a critical point at frequency $\omega_{opt}(|\mathbf{k}| \to 0) = \sqrt{C_{\text{hom}}(\mathbf{k}=\mathbf{0})/I} = \sqrt{\sum_{i=1}^{n} \frac{l_i^2}{2A_b I}(\mathbf{t}_i \cdot \mathbf{K}_i \mathbf{t}_i)}$ that equals the corresponding frequency from the Lagrangian model.

By comparing the submatrices (44) with those (13) and noting that $\cos(\mathbf{k} \cdot \mathbf{x}_i) \simeq 1 - \frac{1}{2}(\mathbf{x}_i \otimes \mathbf{x}_i):(\mathbf{k} \otimes \mathbf{k})$ and $\sin(\mathbf{k} \cdot \mathbf{x}_i) \simeq \mathbf{k} \cdot \mathbf{x}_i$, it follows that $\mathbf{C}_{Lag}(\mathbf{k},\omega) = A_b \mathbf{C}_{\text{hom}}(\mathbf{k},\omega) + \mathcal{O}(|\mathbf{k}|^3)$, therefore the hermitian matrix of the Lagrangian system $\mathbf{C}_{Lag}(\mathbf{k},\omega)$ may be approximated by the corresponding one from the micropolar homogenized model considering the second order expansion of the matrix in the wave vector $\mathbf{k}$. It may be easily verified that if the elastic positive defined second order tensor $\mathbf{E}_s^+$ given by (39) is assumed, that is derived by a first order expansion of the rotation field, the optical branch turns out to be approximated by the equivalent micropolar continuum with a lower accuracy. This circumstance is shown in the examples of Section 5 where the possibility of loss of strong hyperbolicity in cases of wave propagation for the examples considered is investigated.

## 5. Examples

Let us consider now some cases of blocky materials consisting of different periodic planar tessellations in order to understand the influence of both the block shape and the elasticity of the interfaces on the dispersion functions, which are obtained both with the exact Lagrangian model and with the approximate micropolar model. This analysis is also aimed at understanding the limits of validity of the micropolar model. Three cases are analyzed, namely rhombic and hexagonal tessellations and an assemblage of rectangles that is representative of running bond masonry or biological systems like nacre. The acoustic band structures given by the discrete model formulated in Section 2 is represented in the Brillouin zone (see Brillouin, 1953). In particular, this representation is given along the closed polygonal curve $\Upsilon$ with vertices identified by the values $\Xi_j$ ($j=0,..,6$ for the rhombic tiling and running bond tessellation, and $j=0,..,3$ for hexagonal tiling) of the arch-length $\Xi$ in the dimensionless plane $(k_1 l, k_2 l)$ (see

Figures 6.c, 10.c) for rhombic and hexagonal tiling and in the dimensionless plane $(k_1 b, k_2 a)$ (see Figures 14.c) for running bond tessellation. A compact spectral description is given in term of the dimensionless frequency $\omega/\sqrt{k_n l/M}$ (with $l$ the characteristic size of the block) as a function of the arch-length $\Xi$, in terms of the dimensionless geometrical parameter of the block and of the elasticities of the interface. Here the interface is assumed active on the whole length of the connected edge.

*5.1. Periodic rhombic tiling*

In the considered rhombic tessellation shown in Figure 5, the rhomb block has length side $l$, interior angle $\pi/2 - \alpha$, area $A_b = l^2 \cos\alpha$ and mass moment of inertia $J = l^2 M/6$. The vectors representative of the N=2 blocks surrounding the reference block are $\mathbf{n}_1 = -\mathbf{n}_3 = -\mathbf{d}_2^\perp = \mathbf{d}_4^\perp = \mathbf{e}_1$,

$$\mathbf{t}_1 = -\mathbf{t}_3 = \mathbf{d}_2 = -\mathbf{d}_4 = \mathbf{e}_2, \qquad \mathbf{n}_2 = -\mathbf{n}_4 = \mathbf{d}_1^\perp = -\mathbf{d}_3^\perp = \sin\alpha\,\mathbf{e}_1 + \cos\alpha\,\mathbf{e}_2,$$

$\mathbf{t}_2 = -\mathbf{t}_4 = \mathbf{d}_3 = -\mathbf{d}_1 = -\cos\alpha\,\mathbf{e}_1 + \sin\alpha\,\mathbf{e}_2$, with $\beta_1 = -\alpha$ and $\beta_2 = \alpha$. Since $b = l$, the overall elastic parameters of the interfaces are $K_n = K_n^i = k_n l$, $K_t = K_t^i = k_t l$ and $K_\varphi = K_\varphi^i = k_n l^3/12$, so that the eigenvalue problem of the harmonic plane wave propagation in the discrete model derived by the Lagrangian model takes (12) the form:

$$\begin{bmatrix} 2(\bar{K}\mathrm{f}_1 + K_t \mathrm{f}_2) - \omega^2 M & -2\tilde{K}\sin 2\alpha\,\mathrm{f}_1 & -2il(\tilde{K}\sin 2\alpha\,\mathrm{g}_1 + K_t \cos\alpha\,\mathrm{g}_2) \\ -2\tilde{K}\sin 2\alpha\,\mathrm{f}_1 & 2(\hat{K}\mathrm{f}_1 + K_n \mathrm{f}_2) - \omega^2 M & 2il(\hat{K}\,\mathrm{g}_1 + K_n \sin\alpha\,\mathrm{g}_2) \\ 2il(\tilde{K}\sin 2\alpha\,\mathrm{g}_1 + K_t \cos\alpha\,\mathrm{g}_2) & -2il(\hat{K}\,\mathrm{g}_1 + K_n \sin\alpha\,\mathrm{g}_2) & 2K_\varphi(\mathrm{f}_1 + \mathrm{f}_2) + \dfrac{l^2}{2}(K_n \mathrm{f}_1 + \hat{K}\mathrm{f}_2) - \omega^2 J \end{bmatrix} \begin{Bmatrix} \hat{u}_1 \\ \hat{u}_2 \\ \hat{\phi} \end{Bmatrix} = \mathbf{0}$$

(45)

where the following terms are defined $\hat{K} = (K_n \sin^2\alpha + K_t \cos^2\alpha)$, $\bar{K} = (K_n \cos^2\alpha + K_t \sin^2\alpha)$,

$\tilde{K} = \dfrac{(K_n - K_t)}{2}\sin 2\alpha$, $\mathrm{f}_1 = [1 - \cos(k_1 l)]$, $\mathrm{f}_2 = [1 - \cos(\sin\alpha k_1 l + \cos\alpha k_2 l)]$, $\mathrm{g}_1 = \sin(k_1 l)$,

$\mathrm{g}_2 = \sin(\sin\alpha k_1 l + \cos\alpha k_2 l)$.

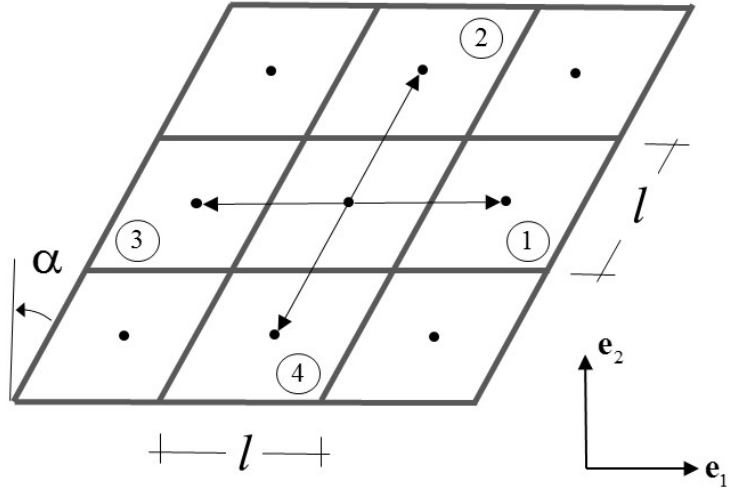

Figure 5. The rhombic tessellation.

In the long wave approximation, the elasticity tensors obtained from (24) and (28) have the following independent elastic moduli in the reference $(\mathbf{e}_1, \mathbf{e}_2)$:

$$E_{1111} = \frac{\bar{K}(1+\sin^4\alpha) - 2\tilde{K}\sin^3\alpha\cos\alpha + \hat{K}\sin^2\alpha\cos^2\alpha}{2\cos\alpha}$$

$$E_{1122} = \frac{\bar{K}\sin^2\alpha\cos^2\alpha - \tilde{K}\sin\alpha\cos\alpha\cos 2\alpha - \hat{K}\sin^2\alpha\cos^2\alpha}{2\cos\alpha}$$

$$E_{1112} = \frac{\bar{K}\sin^3\alpha\cos\alpha - 2\tilde{K}\sin^2\alpha\cos^2\alpha + \hat{K}\sin\alpha\cos^3\alpha}{2\cos\alpha}$$

$$E_{1121} = \frac{\bar{K}\sin^3\alpha\cos\alpha - \tilde{K}(1+\sin^2\alpha\cos^2\alpha - \sin^4\alpha) - \hat{K}\sin^3\alpha\cos\alpha}{2\cos\alpha}$$

$$E_{2222} = \frac{\bar{K}\cos^4\alpha + 2\tilde{K}\sin\alpha\cos^3\alpha + \hat{K}\sin^2\alpha\cos^2\alpha}{2\cos\alpha}$$

$$E_{2212} = \frac{\bar{K}\sin\alpha\cos^3\alpha - \tilde{K}\cos^2\alpha\cos 2\alpha - \hat{K}\sin\alpha\cos^3\alpha}{2\cos\alpha}$$

$$E_{2221} = \frac{\bar{K}\sin\alpha\cos^3\alpha + 2\tilde{K}\sin^2\alpha\cos^2\alpha + \hat{K}\sin^3\alpha\cos\alpha}{2\cos\alpha}$$

$$E_{1212} = \frac{\bar{K}\sin^2\alpha\cos^2\alpha - 2\tilde{K}\sin\alpha\cos^3\alpha + \hat{K}\cos^4\alpha}{2\cos\alpha}$$

$$E_{1221} = \frac{\bar{K}\sin^2\alpha\cos^2\alpha - 2\tilde{K}\sin\alpha\cos\alpha\cos 2\alpha - \hat{K}\sin^2\alpha\cos^2\alpha}{2\cos\alpha} \qquad (46)$$

$$E_{2121} = \frac{\bar{K}\sin^2\alpha\cos^2\alpha + 2\tilde{K}\sin^3\alpha\cos\alpha + \hat{K}(1+\sin^4\alpha)}{2\cos\alpha}$$

$$E_{11} = \left[K_\varphi - \frac{l^2}{4}\hat{K}\right]\frac{(1+\sin^2\alpha)}{2\cos\alpha}$$

$$E_{22} = \left[K_\varphi - \frac{l^2}{4}\hat{K}\right]\frac{\cos\alpha}{2}$$

$$E_{12} = \left[K_\varphi - \frac{l^2}{4}\hat{K}\right]\frac{\sin\alpha}{2}$$

where the symmetric terms are omitted. In case of square shape of the blocks $(\alpha=0)$ it follows $\beta_i = 0$ and the constitutive equation take the simple form in the Voigt notation

$$\begin{Bmatrix}\sigma_{11}\\\sigma_{22}\\\sigma_{12}\\\sigma_{21}\\m_1\\m_2\end{Bmatrix} = \frac{1}{2}\begin{bmatrix}\bar{K} & 0 & 0 & 0 & 0 & 0\\0 & \bar{K} & 0 & 0 & 0 & 0\\0 & 0 & \hat{K} & 0 & 0 & 0\\0 & 0 & 0 & \hat{K} & 0 & 0\\0 & 0 & 0 & 0 & K_\varphi - \frac{l^2}{4}\hat{K} & 0\\0 & 0 & 0 & 0 & 0 & K_\varphi - \frac{l^2}{4}\hat{K}\end{bmatrix}\begin{Bmatrix}\gamma_{11}\\\gamma_{22}\\\gamma_{12}\\\gamma_{21}\\\chi_1\\\chi_2\end{Bmatrix}. \qquad (47)$$

In this case, the second order elastic tensor relating the micro-curvatures to the micro-couples turns out to be positive defined if and only if the constitutive condition is verified $k_n > 12k_t$, i.e. the normal stiffness is more than one order of magnitude of the tangential one. In case of first order expansion of the rotation field, the elastic moduli involving the curvature take the form form equation (39):

$$E_{11}^+ = \left[K_\varphi + \frac{l^2}{4}\hat{K}\right]\frac{(1+\sin^4\alpha)}{2\cos\alpha}, \quad E_{22}^+ = \frac{1}{2}\left[K_\varphi + \frac{l^2}{4}\hat{K}\right]\cos\alpha, \quad E_{12}^+ = \frac{1}{2}\left[K_\varphi + \frac{l^2}{4}\hat{K}\right]\sin\alpha . \qquad (48)$$

In this case the approximate equivalent continuum formulation with the parameters given in (44) provides the secular equation system written as follows:

$$\begin{bmatrix}\frac{\bar{K}k_1^2 + K_t(\sin\alpha\, k_1 + \cos\alpha\, k_2)^2}{\cos\alpha} - \rho\omega^2 & \frac{\tilde{K}k_1^2}{\cos\alpha} & i\left[\frac{2\tilde{K}k_1}{\cos\alpha} - K_t(\sin\alpha\, k_1 + \cos\alpha\, k_2)\right]\\\frac{\tilde{K}k_1^2}{\cos\alpha} & \frac{\hat{K}k_1^2 + K_n(\sin\alpha\, k_1 + \cos\alpha\, k_2)^2}{\cos\alpha} - \rho\omega^2 & i\left[\frac{\tilde{K}k_1}{\cos\alpha} + K_n\tan\alpha(\sin\alpha\, k_1 + \cos\alpha\, k_2)\right]\\-i\left[\frac{2\tilde{K}k_1}{\cos\alpha} - K_t(\sin\alpha\, k_1 + \cos\alpha\, k_2)\right] & -i\left[\frac{\tilde{K}k_1}{\cos\alpha} + K_n\tan\alpha(\sin\alpha\, k_1 + \cos\alpha\, k_2)\right] & \frac{1}{\cos\alpha}\left\{2\hat{K} + \left(K_\varphi - \frac{l^2}{4}\hat{K}\right)\left[k_1^2 + (\sin\alpha\, k_1 + \cos\alpha\, k_2)^2\right]\right\} - I\omega^2\end{bmatrix}\begin{Bmatrix}\hat{u}_1\\\hat{u}_2\\\hat{\phi}\end{Bmatrix} = \mathbf{0}$$

(49)

In the long wavelength limit $\lambda \to \infty$, namely $|\mathbf{k}| = 0$ the acoustic $\omega_{aco1,2} = 0$ and optical $\omega_{opt} = \sqrt{\dfrac{2\hat{K}}{\cos\alpha I_1}}$ frequencies are obtained.

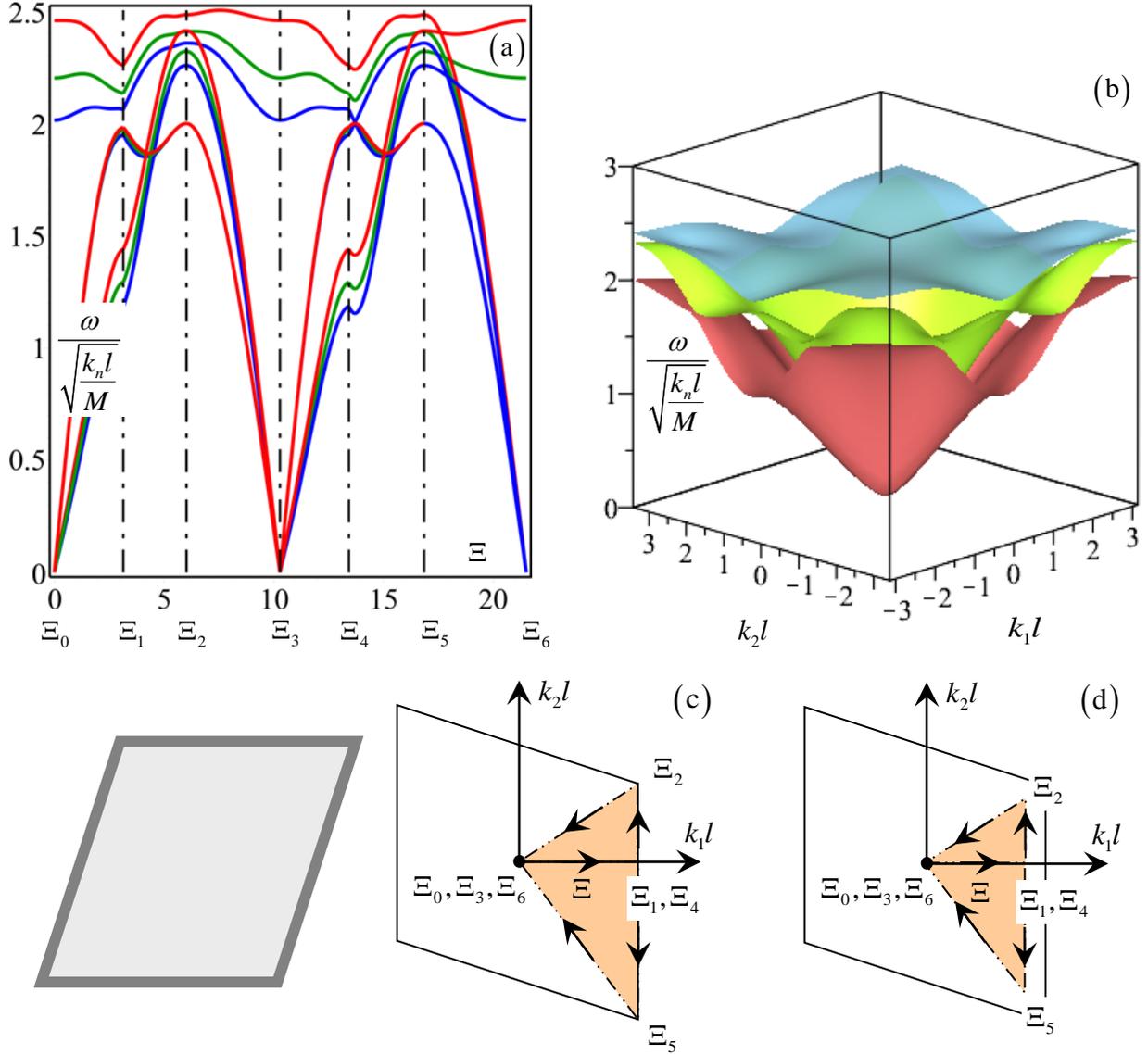

Figure 6: (a) Influence of the interface stiffness ratio $k_t/k_n$ on the band structure of the rhombic tiling along the closed polygonal curve $\Upsilon$ ($k_t/k_n = 1/2$ red; $k_t/k_n = 2/5$ green; $k_t/k_n = 1/3$ blue) for $\alpha = \pi/36$; (b) dispersive surface for $\alpha = \pi/36$ and $k_t/k_n = 2/5$ (c) Periodic cell and Brillouin zone (highlighted in orange the reduced Brillouin zone bounded by the curve $\Upsilon$); (d) Subdomain of the reduced Brillouin zone.

The dispersive functions of the rhombic tiling with $\alpha = \pi/36$ obtained by the discrete model are shown in Figure 6.a for different interfaces stiffness ratios $k_t/k_n$. The dimensionless angular frequency $\omega/\sqrt{k_n l/M}$ is expressed as function of the arch length $\Xi$ measured on the closed polygonal curve $\Upsilon$ with vertices identified by the values $\Xi_j$, $j = 0,..,6$ (see Figure 6.c). Two acoustic branches are shown together with the optical branch, that exhibits a critical point ($v_g = 0$) at $\Xi_0$, namely in the long wavelength regime ($\mathbf{k} = \mathbf{0}$), with a frequency $\omega_{opt}$ increasing with the interface stiffness ratio $k_t/k_n$. The acoustic spectra exhibit several point of crossing between the optical branch and the acoustical ones. For the stiffness ratios $k_t/k_n = 1/2$ and $k_t/k_n = 2/5$, in the first and fourth segment of the polygonal curve $\Upsilon$, veering phenomena are observed between the optical and the second acoustical one, i.e. a repulsion between the two branches. Finally, it is worth to note the absence of stop band in the band structure.

A more complete description of the acoustic characteristics of the blocky material is given by the dispersion surfaces in the Brillouin zone (see Figure 6.c) shown in Figure 6.b, representing the dimensionless frequency $\omega/\sqrt{k_n l/M}$ in terms of the non-dimensional components of the wave vector $(k_1 l, k_2 l)$ for $\alpha = \pi/36$ and $k_t/k_n = 2/5$. In the domain of considered wave vectors, a high spectral density is observed because the optical surface intersects the higher acoustic one, which exhibits, in different regions, a higher frequency.

The accuracy of the micropolar model derived in Section 3 is here analysed by comparing the band structure of the discrete model with that one by the equivalent homogeneous Cosserat model characterized by the overall constitutive tensors $\mathbb{E}_s$ and $\mathbf{E}_s$ whose components are given in equation (46). This comparison is carried out in a homotetic subdomain of the reduced Brillouin zone bounded by the closed polygonal curve $\Upsilon$ (see Figure 6.d) and is shown by the diagrams of Figure 7.a. In consideration of the property of the hermitian matrices of the systems under consideration $\mathbf{C}_{Hom}(\mathbf{k},\omega) = \mathbf{C}_{Lag}(\mathbf{k},\omega) + O(|\mathbf{k}|^3)$, it results that the difference in frequency between the discrete and continuum model is lower than 5% for dimensionless wave numbers $|\mathbf{k}|l \leq 1.6$, i.e. wavelengths $\lambda \geq 3.94 l$. The same comparison is carried out with reference to the micropolar continuum model characterized by the overall positive defined second order tensor $\mathbf{E}_s^+$, whose components are given in (48). The comparison of the band structure from this model with that from the discrete model is represented in Figure 7.b, where a low accuracy of this micropolar model come out in describing the optical branch, while a good agreement is obtained for the acoustic branches.

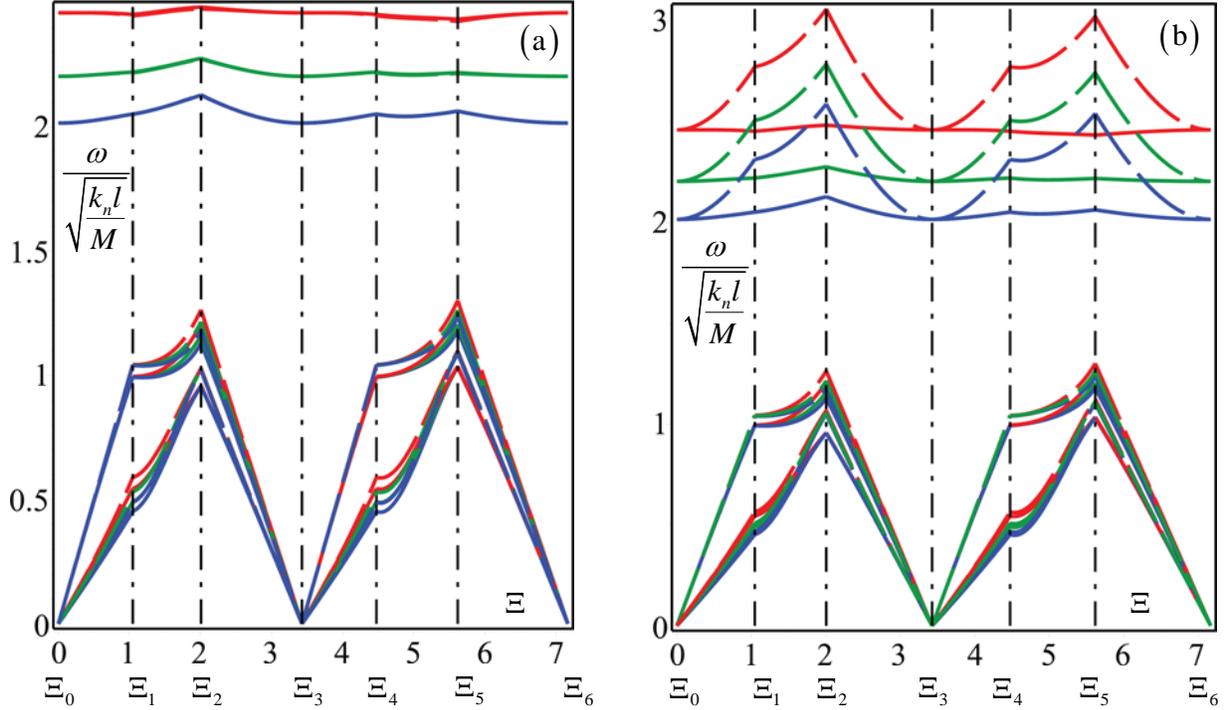

Figure 7: Dispersive functions for rhombic tiling ($\alpha = \pi/36$). Comparison between the discrete model (continuous line) and the micropolar continuum model (dashed line) in a subdomain of the reduced Brillouin zone ($k_t/k_n = 1/2$ red; $k_t/k_n = 2/5$ green; $k_t/k_n = 1/3$ blue). (a) Constitutive tensor $\mathbf{E}_s$; (b) Constitutive tensor $\mathbf{E}_s^+$.

The dispersion surfaces in the Brillouin zone obtained by assuming the elastic moduli (46) are shown in Figure 8.a in terms of $\omega^2 M/k_n l$ and of the components of the dimensionless wave vector $(k_1 l, k_2 l)$ for $\alpha = \pi/36$ and $k_t/k_n = 2/5$. Although the positive definiteness of the second order tensor $\mathbf{E}_s$ is not guarantee, the acoustic and optical surfaces (red, green and cyan surfaces) are always positive in the Brillouin zone and consequently the phase velocities are real, according to the Legendre-Hadamard ellipticity condition. In Figure 8.b, the lower acoustic surface is represented in terms of in the Brillouin zone, where the phase velocity comes out to be real in all points of the considered domain (in red) and hence also in the subdomain $-\pi/2 \leq k_i l \leq \pi/2$, with $i = 1,2$, where the acoustic response of the micropolar continuum model is in good agreement with that of the discrete Lagrangian model, with an error less than 10%.

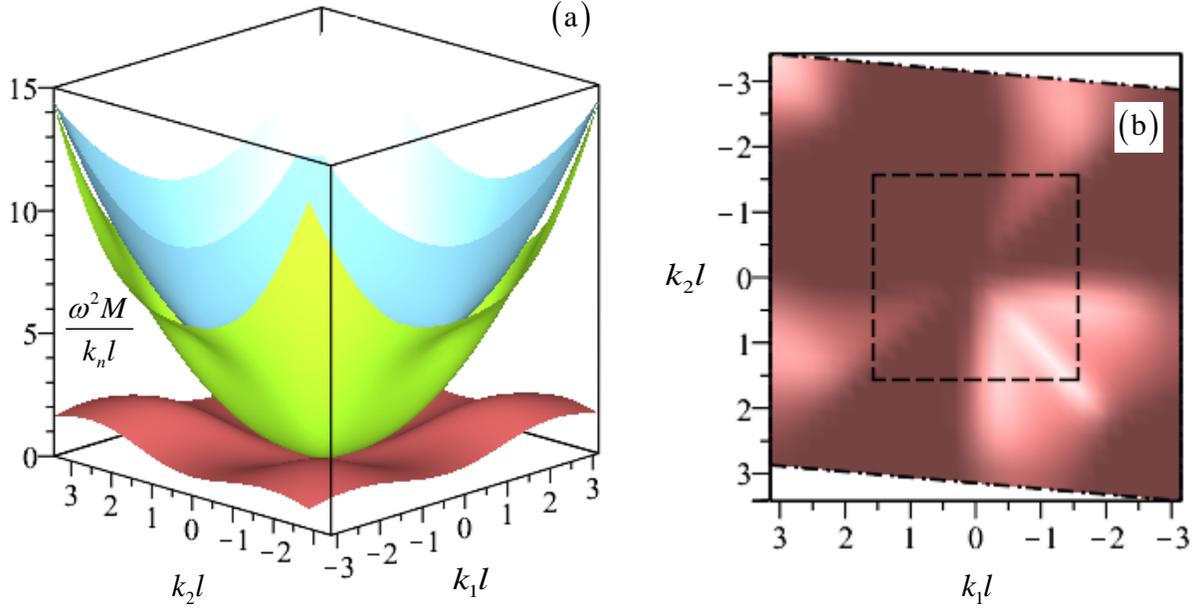

Figure 8: Dispersive surfaces in terms of $\omega^2 M/k_n l$ in the Brillouin zone obtained by micropolar continuum model with constitutive tensor $\mathbf{E}_s$ for rhombic tiling ($\alpha = \pi/36$, $k_t/k_n = 2/5$).
(a) Positive Floquet-Bloch spectrum; (b) Domain of positivity of the first acoustic surface and square subdomain of good accuracy of the model.

*5.2. Hexagonal tiling*

The hexagonal tessellation is shown in Figure 5, with the hexagonal block having an apothem size $l/2$, area $A_b = \frac{\sqrt{3}}{2} l^2$ and mass moment of inertia $J = \frac{5}{12} l^2 M$. The vectors representative of the $N=3$ blocks surrounding the reference block are $\mathbf{n}_1 = \mathbf{e}_1$, $\mathbf{t}_1 = \mathbf{e}_2$, $\mathbf{n}_2 = \frac{1}{2}\mathbf{e}_1 + \frac{\sqrt{3}}{2}\mathbf{e}_2$, $\mathbf{t}_2 = -\frac{\sqrt{3}}{2}\mathbf{e}_1 + \frac{1}{2}\mathbf{e}_2$, $\mathbf{n}_3 = -\frac{1}{2}\mathbf{e}_1 + \frac{\sqrt{3}}{2}\mathbf{e}_2$, $\mathbf{t}_3 = -\frac{\sqrt{3}}{2}\mathbf{e}_1 - \frac{1}{2}\mathbf{e}_2$ and $\beta_i = 0$, $i=1,3$.

The overall elastic parameters of the interfaces are $K_n = K_n^i = \sqrt{3}/3 \, k_n l$, $K_t = K_t^i = \sqrt{3}/3 \, k_t l$ and $K_\varphi = K_\varphi^i = \sqrt{3}/108 \, k_n l^3$.

The eigenvalue problem of the harmonic plane wave propagation in the discrete model is:

$$\begin{bmatrix} \frac{1}{2}\left[(4f_1+f_2+f_3)K_n+3(f_2+f_3)K_t\right]-\omega^2 M & \frac{\sqrt{3}}{2}\left[(K_n-K_t)(f_2-f_3)\right] & -i\frac{\sqrt{3}}{2}(g_2+g_3)K_t l \\ \frac{\sqrt{3}}{2}\left[(K_n-K_t)(f_2-f_3)\right] & \frac{1}{2}\left[3(f_2+f_3)K_n+(4f_1+f_2+f_3)K_t\right]-\omega^2 M & i\frac{1}{2}(2g_1+g_2-g_3)K_t l \\ i\frac{\sqrt{3}}{2}(g_2+g_3)K_t l & -i\frac{1}{2}(2g_1+g_2-g_3)K_t l & 2\left[K_\varphi(f_1+f_2+f_3)+\frac{l^2}{2}K_t(6-f_1-f_2-f_3)\right]-\omega^2 J \end{bmatrix}\begin{Bmatrix}\hat{u}_1\\\hat{u}_2\\\hat{\varphi}\end{Bmatrix}=\mathbf{0}$$

(50)

where $f_i = f_i(\mathbf{k},\mathbf{n}_i) = 1-\cos(l\mathbf{k}\cdot\mathbf{n}_i)$ and $g_i = g_i(\mathbf{k},\mathbf{n}_i) = \sin(l\mathbf{k}\cdot\mathbf{n}_i)$, $i=1,3$.

In the long wave approximation based on the equivalent continuum, the constitutive equation is obtained from (24) and (28) and is written as follows

$$\begin{Bmatrix}\sigma_{11}\\\sigma_{22}\\\sigma_{12}\\\sigma_{21}\\m_1\\m_2\end{Bmatrix}=\begin{bmatrix}2\mu+\lambda & \lambda & 0 & 0 & 0 & 0\\\lambda & 2\mu+\lambda & 0 & 0 & 0 & 0\\0 & 0 & \mu+k & \mu-k & 0 & 0\\0 & 0 & \mu-k & \mu+k & 0 & 0\\0 & 0 & 0 & 0 & S & 0\\0 & 0 & 0 & 0 & 0 & S\end{bmatrix}\begin{Bmatrix}\gamma_{11}\\\gamma_{22}\\\gamma_{12}\\\gamma_{21}\\\chi_1\\\chi_2\end{Bmatrix},\qquad(51)$$

in which the four elastic moduli

$$\mu=\frac{\sqrt{3}}{4}(K_n+K_t),\quad \lambda=\frac{\sqrt{3}}{4}(K_n-K_t),\quad \kappa=\frac{\sqrt{3}}{2}K_t,\quad S=\sqrt{3}\left[K_\varphi-\frac{l^2}{4}K_t\right],\qquad(52)$$

depend on the interface stiffnesses and on the characteristic length $l$ of the lattice ($\lambda$, $\mu$ Lamé constants, $\kappa$ and $S$ additional micropolar elastic coefficients). In case of simplified first order expansion of the rotational field the elastic modulus involving the curvature is obtained $S^+ = \sqrt{3}\left[K_\varphi+\frac{l^2}{4}K_t\right]$. In case of symmetric macro-strain fields, the fourth order elastic tensor for the hexagonal system corresponds to that of the transversely isotropic system whose elastic moduli in the plane of the lattice are:

$$E_{\text{hom}} = 2\sqrt{3}\frac{K_n(K_n+K_t)}{3K_n+K_t},\qquad \nu_{\text{hom}} = \frac{K_n-K_t}{3K_n+K_t},\qquad(53)$$

i.e. the resulting equivalent classical continuum turns out to be auxetic (Prawoto, 2012) in case the interface elastic moduli were $K_t > K_n$. This condition has been obtained and discussed by Shufrin *et al.*, 2012, who, instead of considering the linear interface, he proposed a connecting element between the opposite sides of two adjacent hexagons, which is constituted by two opposite arches having a greater transverse stiffness than the corresponding normal, namely $K_t > K_n$.

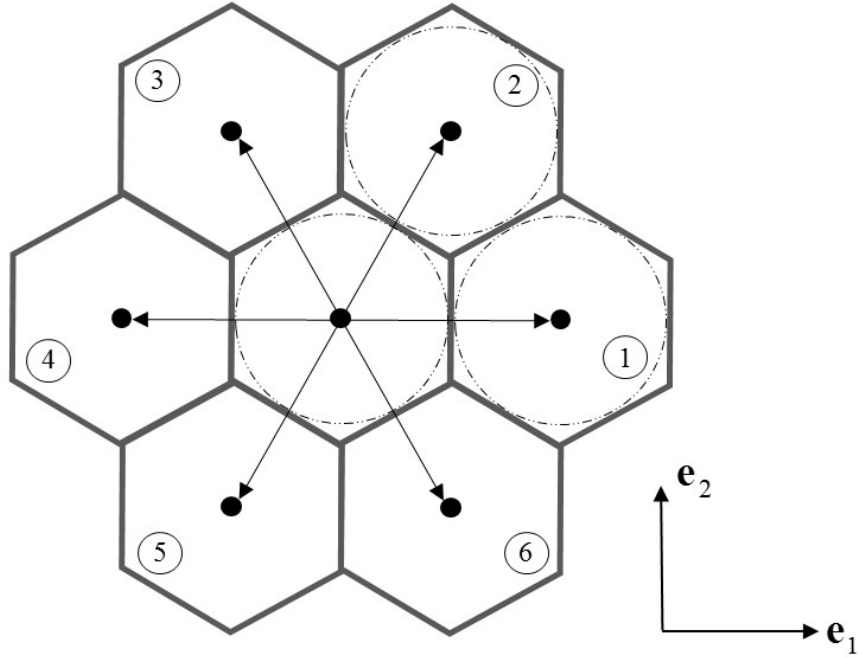

Figure 9: Hexagonal tessellation (in dashed double dot line the limit case of granular packed hexagonal structure).

The equation providing the dispersion function is written in the standard form of plane micropolar centro-symmetric models

$$\begin{bmatrix} (2\mu+\lambda)k_1^2 + (\mu+\kappa)k_2^2 - \rho\omega^2 & (\mu+\lambda-\kappa)k_1 k_2 & -2i\kappa k_2 \\ (\mu+\lambda-\kappa)k_1 k_2 & (\mu+\kappa)k_1^2 + (2\mu+\lambda)k_2^2 - \rho\omega^2 & 2i\kappa k_1 \\ 2i\kappa k_2 & -2i\kappa k_1 & S(k_1^2+k_2^2)+4\kappa - I\omega^2 \end{bmatrix} \begin{Bmatrix} \hat{u}_1 \\ \hat{u}_2 \\ \hat{\phi} \end{Bmatrix} = \mathbf{0} \ .$$

(54)

For long wavelength propagation $\lambda \to \infty$, namely **k=0**, the frequencies are $\omega_{aco1,2} = 0$ and

$$\omega_{opt} = 2\sqrt{\frac{\kappa}{I}} = \sqrt[4]{3}\sqrt{\frac{2K_t}{I}}.$$

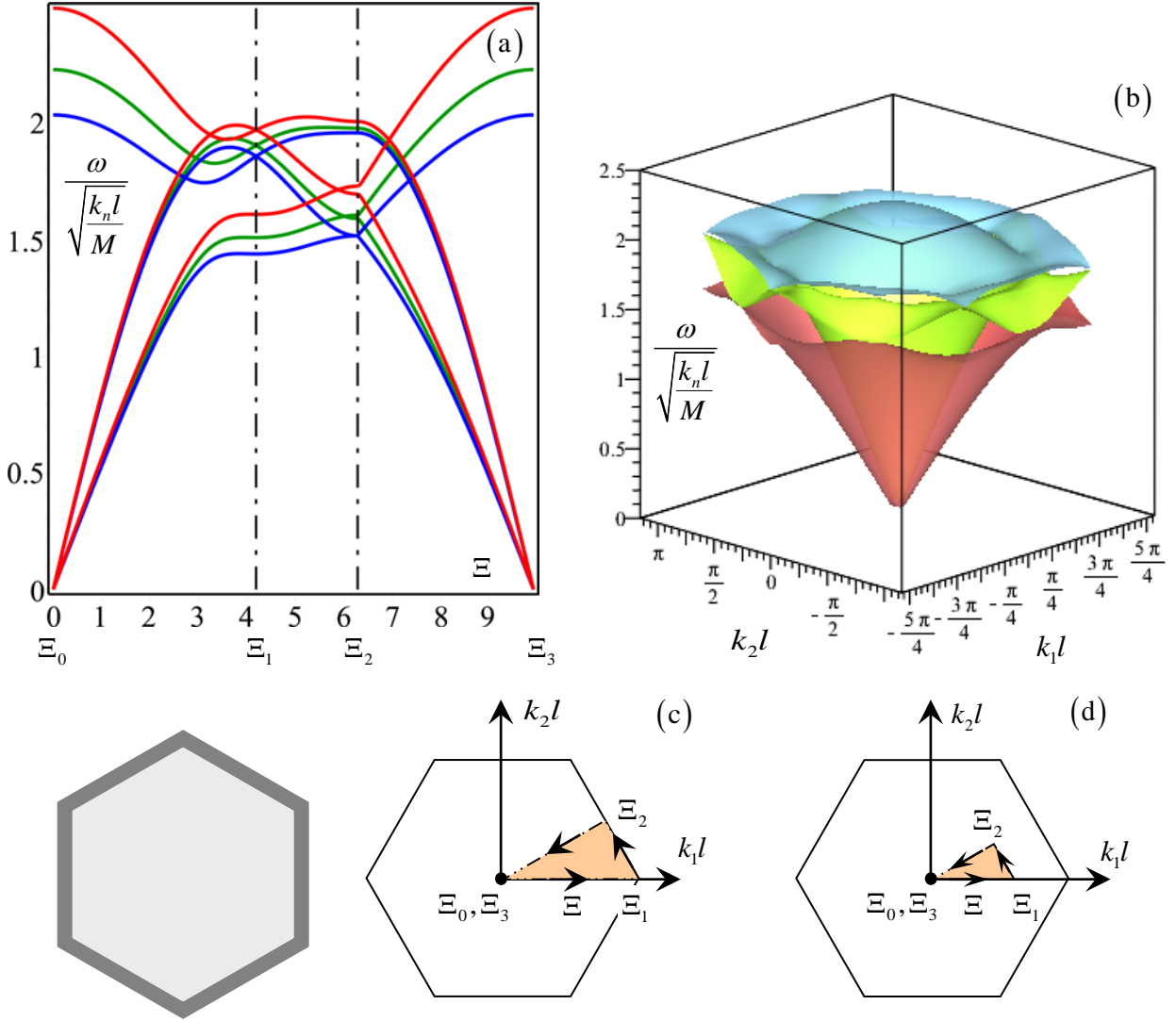

Figure 10: (a) Influence of the interface stiffness ratio $k_t/k_n$ on the band structure of the hexahonal tiling along the closed polygonal curve $\Upsilon$ ($k_t/k_n = 1/2$ red; $k_t/k_n = 2/5$ green; $k_t/k_n = 1/3$ blue); (b) dispersive surface for $k_t/k_n = 2/5$ (c) Periodic cell and Brillouin zone (highlighted in orange the irreducible Brillouin zone bounded by the curve $\Upsilon$); (d) Subdomain of the irreducible Brillouin zone.

For the limit case of a hexagonal packed structure of circular blocks (see Figure 9) with punctual contact among them, which is representative of a regular granular material, the interface is characterized by a limit extension $b_i \to 0$ and by a stiffness ratio $K_\varphi/K_{n,s} = \mathcal{O}(b^2)$. Consequently, the rotational stiffness $K_\varphi$ turns out to be negligible in equations (50), (52) and (54), and the optical branch in the Lagrangian model turns out to be decreasing with the wave number increasing in the neighbourhood of **k**=**0**. The same acoustic behaviour is obtained through the homogeneous micropolar model by solving the eigenvalue problem (54), being $S < 0$ (see the fourth term in equation (52)). This outcome gives a response to the problem raised by Merkel *et al.*, 2011, who criticized the Cosserat model (characterized by $S > 0$) for its inability to simulate the downward concavity of the optical branch in the neighbourhood of **k**=**0** obtained in the Lagrangian model.

The dispersive functions are given in Figure 10.a for different values of the stiffness ratio $k_t/k_n$ in terms of the dimensionless frequency $\omega/\sqrt{k_n l/M}$ as function of the arch length $\Xi$ on the closed polygonal curve $\Upsilon$ with vertices identified by the values $\Xi_j$, $j = 0,..,3$, (see Figure 10.c). In agreement with the previous case, the frequency $\omega_{opt}$ of the critical point on the optical branch in the long wavelength regime is increasing with the ratio $k_t/k_n$. Several crossing points may be observed while no stop band takes place. In Figure 10.b the dispersive surfaces are shown for $k_t/k_n = 2/5$ in terms of the non-dimensional components of the wave vector $(k_1 l, k_2 l)$, together with a high spectral density as already noted for the previous case.

The comparison between the band structure of the discrete model and the corresponding one from the equivalent homogeneous Cosserat model (with elastic constant given in (52) is represented in the diagrams of Figure 11.a. This comparison is carried out in a homotetic subdomain of the reduced Brillouin zone bounded by the closed polygonal curve $\Upsilon$ (see Figure 10.d). Also in this case the micropolar model appears to be very accurate in the considered domain with an error lower than 5% for dimensionless wave numbers $|\mathbf{k}|l \leq 1.7$ i.e. wavelengths $\lambda \geq 3.75 l$. If the micropolar model with a positive defined elastic modulus $S^+$ is assumed the resulting band structure is shown in Figure 11.b (dashed line) showing again a lower accuracy in simulating the optical branches of the discrete Lagrangian model, while a good accuracy is obtained for the acoustic branches. In particular, it should be noted that with the elastic moduli (52) the downward concavity of the optical branch in the neighborhood of the long wavelength regime may be simulated accurately, while assuming the positive modulus $S^+$ only the upward concavity is possible.

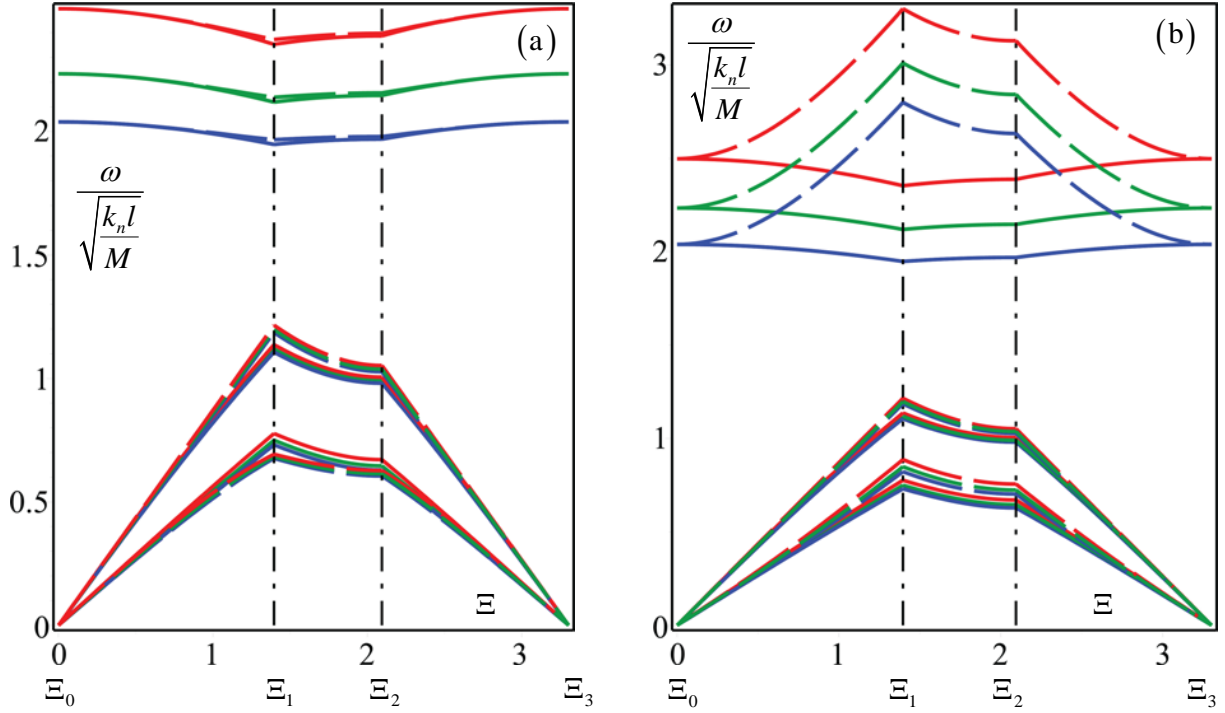

Figure 11: Dispersive functions for hexagonal tiling. Comparison between the discrete model (continuous line) and the micropolar continuum model (dashed line) in a subdomain of the reduced Brillouin zone ($k_t/k_n = 1/2$ red; $k_t/k_n = 2/5$ green; $k_t/k_n = 1/3$ blue).
(a) Constitutive tensor $\mathbf{E}_s$; (b) Constitutive tensor $\mathbf{E}_s^+$.

The dispersion surfaces in the Brillouin zone (see Figure 10.c) obtained by assuming the elastic moduli (52) for the micropolar continuum are given in terms of $\omega^2 M/k_n l$ and of the components of the dimensionless wave vector $(k_1 l, k_2 l)$ for $k_t/k_n = 2/5$ in Figure 12.a. In these diagrams only the positive portions of the surfaces are represented, which correspond to consider the case of real phase velocity. While the higher acoustical surface and the optical one are positive defined in the Brillouin zone, the first (lower) acoustic surface results negative in a narrow region $|\mathbf{k}|l \geq 2\sqrt{2}/3\,\pi$. Nevertheless, it must be observed from the previous considerations that the micropolar model provides good simulations of the acoustic behaviour of the discrete model in the smaller domain $-\pi/2 \leq k_i l \leq \pi/2$, $i=1,2$, shown in the domain of positivity in Figure 12.b. On the basis of this consideration and for the considered case, one may derive that the Legendre-Hadamard condition is less strict than the validity of the simulation of the discrete model.

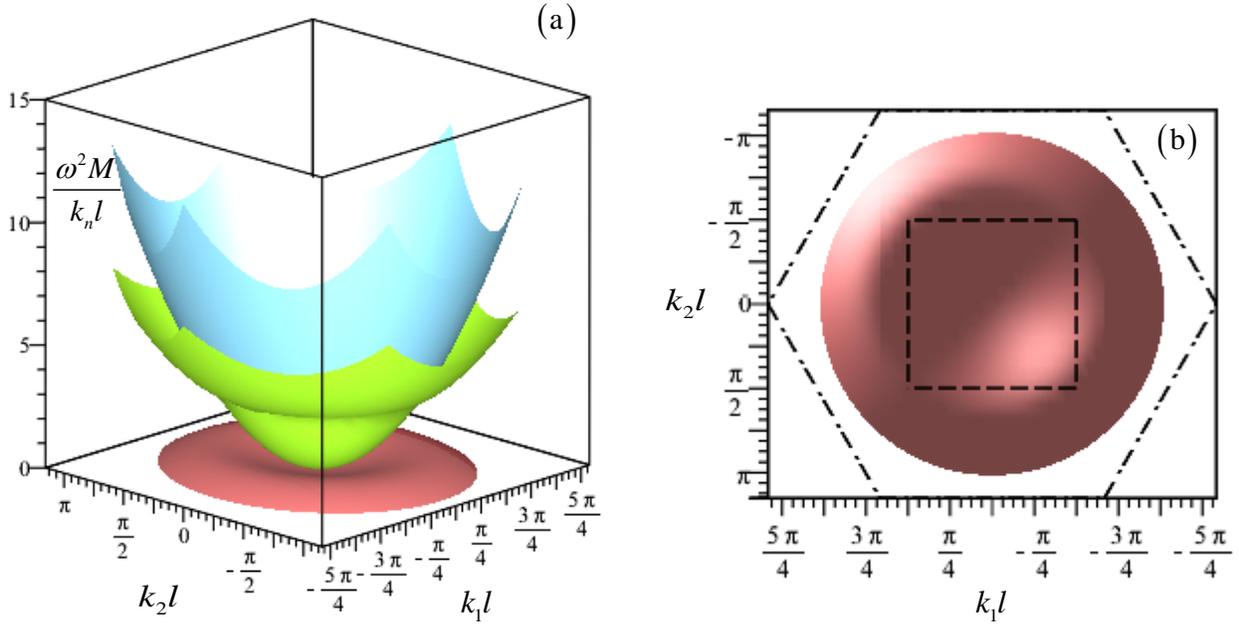

Figure 12: Dispersive surfaces in terms of $\omega^2 M / k_n l$ in the Brillouin zone obtained by micropolar continuum model with constitutive tensor $\mathbf{E}_s$ for hexagonal tiling ($k_t/k_n = 2/5$). (a) Positive Floquet-Bloch spectrum; (b) Domain of positivity of the first acoustic surface and square subdomain of good accuracy of the model.

*5.3. Running bond masonry / Nacre model*

If the regular hexagon considered in the previous case turns into a rectangle with six interfaces, the resulting assembly of blocks is shown in Figure 13 and may be representative of running bond masonry (see Masiani *et al.*, 1995, Sulem J and Mühlhaus, 1997, Trovalusci and Masiani , 2003, 2005, Stefanou *et al.*, 2008, 2010, Pau and Trovalusci, 2012, Baraldi *et al.*, 2015) made of rigid blocks (stone) with elastic interfaces (mortar) or biological and bio-inspired composites, such as the nacre (see for reference Bertoldi *et al.*, 2008, Chen and Wang, 2014, 2015). The rectangular block have width $b$ and height $a$ (with characteristic size $l = \frac{a+b}{2}$), the area $A_b = ab$ and the mass moment of inertia $J = \frac{1}{12}(a^2 + b^2)M$. The vectors and geometrical parameters representative of the $N=3$ blocks surrounding the reference block are $\mathbf{n}_1 = \mathbf{d}_1 = \mathbf{e}_1$, $\mathbf{t}_1 = \mathbf{d}_1^\perp = \mathbf{e}_2$, $\beta_1 = 0$, $l_1 = b$, $\mathbf{n}_2 = \cos\alpha \mathbf{e}_1 + \sin\alpha \mathbf{e}_2$, $\mathbf{t}_2 = -\sin\alpha \mathbf{e}_1 + \cos\alpha \mathbf{e}_2$, $\mathbf{d}_2 = \mathbf{e}_2$, $\mathbf{d}_2^\perp = -\mathbf{e}_1$, $\beta_2 = \frac{\pi}{2} - \alpha$, $l_2 = \sqrt{\left(\frac{b}{2}\right)^2 + a^2}$, $\mathbf{n}_3 = -\cos\alpha \mathbf{e}_1 + \sin\alpha \mathbf{e}_2$ $\mathbf{t}_3 = -\sin\alpha \mathbf{e}_1 - \cos\alpha \mathbf{e}_2$, $\mathbf{d}_3 = \mathbf{e}_2$, $\mathbf{d}_3^\perp = -\mathbf{e}_1$, $\beta_3 = -\beta_2$, $l_3 = l_2$, with $\tan\alpha = 2a/b$. The overall elastic parameters of the interfaces are $K_n^1 = k_n a$, $K_t^1 = k_t a$ and $K_\varphi^1 = k_n a^3/12$, $K_n^{2,3} = \frac{b}{2a} K_n^1$, $K_t^{2,3} = \frac{b}{2a} K_t^1$, $K_\varphi^{2,3} = \left(\frac{b}{2a}\right)^3 K_\varphi^1$.

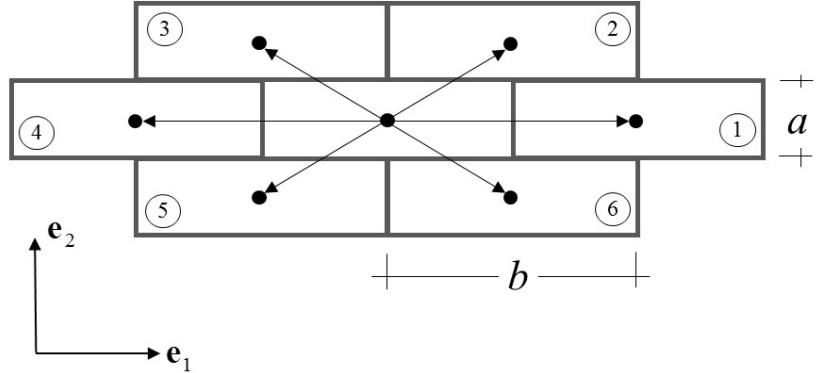

Figure 13: Brick/Nacre tessellation.

The eigenvalue problem of the harmonic plane wave propagation in the discrete model takes the form:

$$\begin{bmatrix} 2\left[K_n^1 f_1 + (f_2 + f_3)K_t^2\right] - \omega^2 M & 0 & -2il_2 \sin\alpha K_t^2 (g_2 + g_3) \\ 0 & 2\left[K_t^1 f_1 + (f_2 + f_3)K_n^2\right] - \omega^2 M & 2i\left[l_1 g_1 K_t^1 + l_2 \cos\alpha K_n^2 (g_2 - g_3)\right] \\ 2il_2 \sin\alpha K_t^2 (g_2 + g_3) & -2i\left[l_1 g_1 K_t^1 + l_2 \cos\alpha K_n^2 (g_2 - g_3)\right] & 2\begin{bmatrix} K_\varphi^1 f_1 + K_\varphi^2 (f_2 + f_3) + \\ + \frac{l_1^2}{4} K_t^1 (2 - f_1) + \frac{l_2^2}{4} K_t^2 (4 - f_2 - f_3) \end{bmatrix} - \omega^2 J \end{bmatrix} \begin{Bmatrix} \hat{u}_1 \\ \hat{u}_2 \\ \hat{\varphi} \end{Bmatrix} = \mathbf{0}$$

(55)

In the long wave approximation based on the equivalent micropolar continuum, the constitutive equation is obtained from (24) and (28) and is written as follows

$$\begin{Bmatrix} \sigma_{11} \\ \sigma_{22} \\ \sigma_{12} \\ \sigma_{21} \\ m_1 \\ m_2 \end{Bmatrix} = \begin{bmatrix} E_{1111} & 0 & 0 & 0 & 0 & 0 \\ 0 & E_{2222} & 0 & 0 & 0 & 0 \\ 0 & 0 & E_{1212} & 0 & 0 & 0 \\ 0 & 0 & 0 & E_{2121} & 0 & 0 \\ 0 & 0 & 0 & 0 & E_{11} & 0 \\ 0 & 0 & 0 & 0 & 0 & E_{22} \end{bmatrix} \begin{Bmatrix} \gamma_{11} \\ \gamma_{22} \\ \gamma_{12} \\ \gamma_{21} \\ \chi_1 \\ \chi_2 \end{Bmatrix}, \qquad (56)$$

in which the four elastic moduli of the fourth order elastic tensor and the two moduli of the second order elastic tensor are

$$E_{1111} = k_n b + k_t \frac{b^2}{4a}, \quad E_{2222} = k_n a \quad E_{1212} = k_t a, \quad E_{2121} = k_n \frac{b^2}{4a} + k_t b,$$

$$E_{11} = \frac{1}{96} k_n a^2 b \left[ 8 - \left( \frac{b^3}{a^3} \right) \right] - \frac{1}{16} k_t ab^2 \left( 1 + 4\frac{b}{a} \right), \quad E_{22} = -\frac{5}{48} k_n ab^2 - \frac{1}{4} k_t a^3, \qquad (57)$$

respectively. The component of the fourth order tensor coincide with those obtained by various researchers for the model running bond masonry (see Sulem *et al.*, 2008, Baraldi *et al.*, 2016) for a comparison). Conversely, the second order tensor components $E_{ij}$ given by equation (28) markedly differ from those available in the literature. Note that for $b = 2a$, the elastic moduli relating the curvatures to the micro-couples are negative defined $E_{11} = -\frac{9}{4} k_t a^3$ and $E_{22} = -\frac{5}{96} k_n b^3 - \frac{1}{32} k_t b^3$. If only a first order expansion is considered for the rotational field, equation (39) applies and the resulting positive defined elasticity constants are

$$E_{11}^+ = \frac{1}{12} k_n a^2 b \left[ 4 + \left( \frac{b}{a} \right)^3 \right] + \frac{1}{16} k_t ab^2 \left( 1 + 4\frac{b}{a} \right), \quad E_{22}^+ = \frac{7}{48} k_n ab^2 + \frac{1}{4} k_t a^3. \qquad (58)$$

Finally, the dispersion funtions for the micropolar equivalent continuum are obtained y solving the following eigenproblem

$$\begin{bmatrix} E_{1111}k_1^2 + E_{1212}k_2^2 - \rho\omega^2 & 0 & -iE_{1212}k_2 \\ 0 & E_{2222}k_1^2 + E_{2121}k_2^2 - \rho\omega^2 & iE_{2121}k_1 \\ iE_{1212}k_2 & -iE_{2121}k_1 & E_{11}k_1^2 + E_{22}k_2^2 + E_{1212} + E_{2121} - I\omega^2 \end{bmatrix} \begin{Bmatrix} \hat{u}_1 \\ \hat{u}_2 \\ \hat{\phi} \end{Bmatrix} = \mathbf{0} \ . \quad (59)$$

For long wavelength limit $\lambda \to \infty$, the frequencies are $\omega_{aco1,2} = 0$ and

$$\omega_{opt} = \sqrt{\frac{E_{1212} + E_{2121}}{I}} = \sqrt{\left[k_n \frac{b^2}{4a} + k_t(a+b)\right]/I} \ .$$

The dispersive functions of the discrete blocky system corresponding to running bond tessellation with $b/a = 2$ are given in the diagrams of Figure 14.a in terms of the dimensionless frequency $\omega/\sqrt{k_n(a+b)/2M}$ as function of the arch length $\Xi$ on the close polygonal curve $\Upsilon$ with vertices $\Xi_j$, $j = 0,..,6$, (see Figure 14.c). Also in this case, the diagrams are obtained for different values of the interface stiffness ratio $k_t/k_n$. In agreement with the previous cases, the frequency $\omega_{opt}$ of the critical point of the optical branch in the long wavelength regime is increasing with the ratio $k_t/k_n$. Several crossing points among the optical and the acoustical branches may be observed together with veering phenomena between the optical branch and the higher acoustical branch, while no band gap is detected. The dispersive surfaces in the Brillouin zone (see Figure 14.c) obtained by the discrete Lagrangian model are represented in Figure 14.b in terms of the dimensionless frequency $\omega/\sqrt{k_n(a+b)/2M}$ in terms of the non-dimensional components of the wave vector $(k_1b, k_2a)$ for $b/a = 2$ and $k_t/k_n = 2/5$. From this diagram, a high spectral density is clearly shown as a result of the interaction of the acoustical and optical branches.

The accuracy of the micropolar model derived in Section 3 is here evaluated for the case of running bond masonry with $b/a = 2$. The comparison between the band structure of the discrete and the micropolar continuum model with elastic moduli given by (57) is shown in the the diagrams of Figure 15.a in a homotetic domain of the reduced Brillouin zone bounded by the closed polygonal curve $\Upsilon$ (see Figure 14.d). Also in this case the micropolar model appears to be very accurate in the considered domain with an error lower than 5% for dimensionless wave numbers $a\left|[2(\mathbf{e}_1 \otimes \mathbf{e}_1) + (\mathbf{e}_2 \otimes \mathbf{e}_2)]\mathbf{k}\right| \leq 2.2$ i.e. wavelengths $\lambda \geq 3.58a$. If the micropolar model

were assumed with elastic moduli given by (58), namely with positive defined second order tensor $\mathbf{E}_s^+$, a good simulation of the acoustic branches from the discrete model would be obtained. On the other side, as shown in the diagrams of Figure 15.b, the optical branch would not properly represented because the different curvature of the function: negative in the discrete model, positive in the micropolar model.

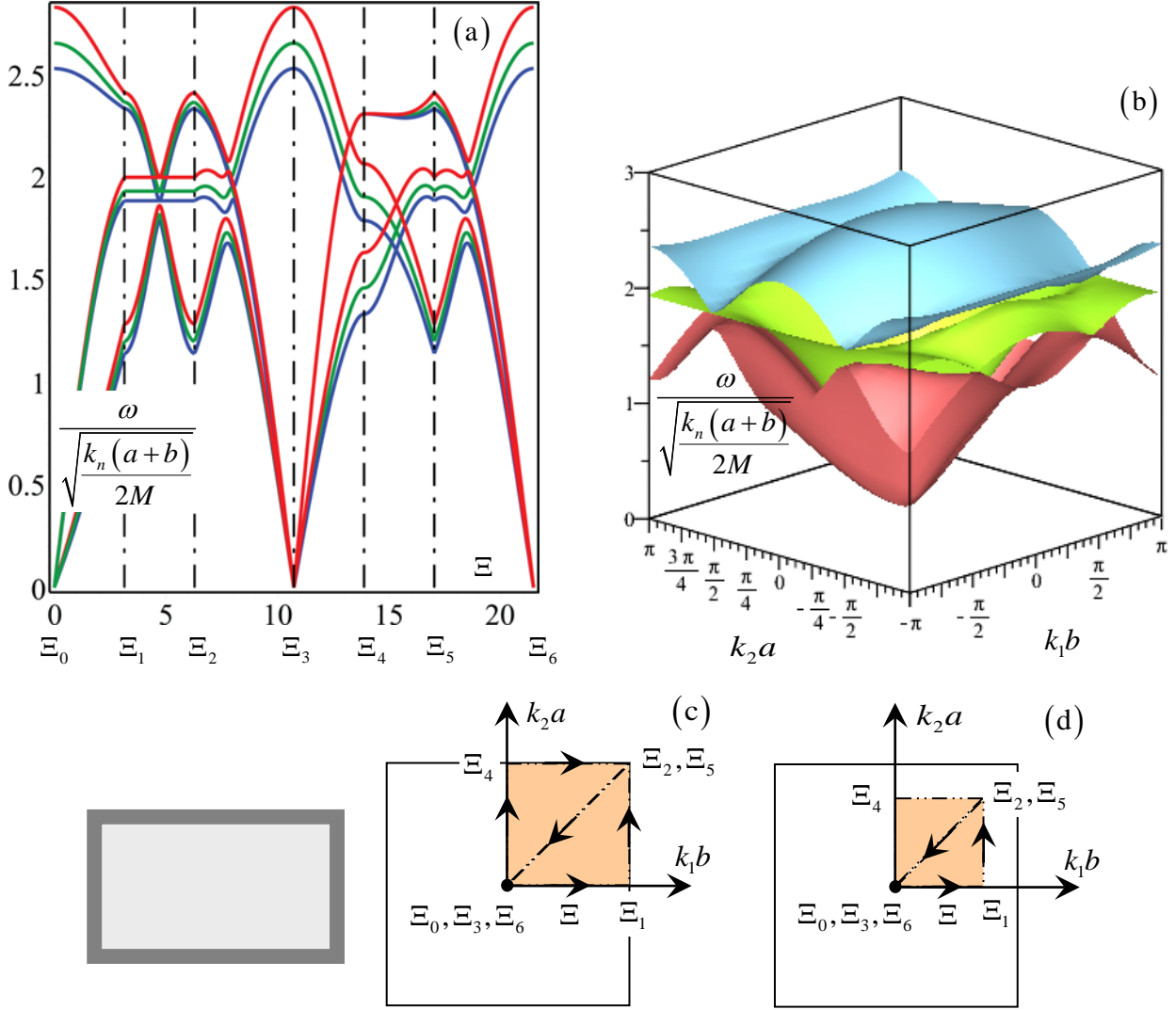

Figure 14: (a) Influence of the interface stiffness ratio $k_t/k_n$ on the band structure of the running bond tessellation along the closed polygonal curve $\Upsilon$ ($k_t/k_n = 1/2$ red; $k_t/k_n = 2/5$ green; $k_t/k_n = 1/3$ blue); (b) dispersive surface for $b/a = 2$ and $k_t/k_n = 2/5$ (c) Periodic cell and Brillouin zone (highlighted in orange the reduced Brillouin zone bounded by the curve $\Upsilon$); (d) Subdomain of the reduced Brillouin zone.

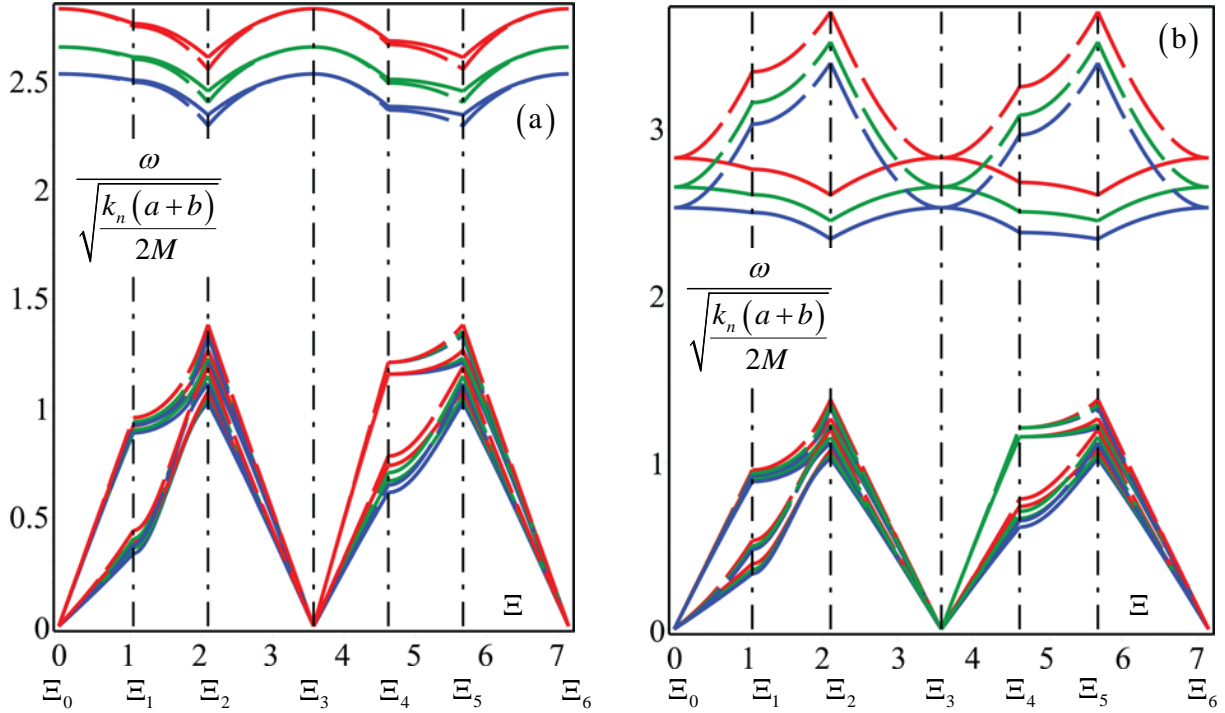

Figure 15: Dispersive functions for running bond tessellation ($b/a = 2$). Comparison between the discrete model (continuous line) and the micropolar continuum model (dashed line) in a subdomain of the reduced Brillouin zone ($k_t/k_n = 1/2$ red; $k_t/k_n = 2/5$ green; $k_t/k_n = 1/3$ blue). (a) Constitutive tensor $\mathbf{E}_s$; (b) Constitutive tensor $\mathbf{E}_s^+$.

The dispersion surfaces in the Brillouin zone (see Figure 14.c) obtained by assuming the elastic moduli (52) for the micropolar continuum are given in terms of $2\omega^2 M / k_n (a+b)$ and of the components of the dimensionless wave vector $(k_1 b, k_2 a)$ for per $b/a = 2$ and $k_t/k_n = 2/5$ in Figure 16.a. As discussed in the previous examples, in these diagrams only the positive portions of the surfaces are represented, which correspond to consider the case of real phase velocity. While the higher acoustical surface and the optical one are positive defined in the Brillouin zone, the first (lower) acoustic surface results negative in a narrow region $a\left|\left[2(\mathbf{e}_1 \otimes \mathbf{e}_1) + (\mathbf{e}_2 \otimes \mathbf{e}_2)\right]\mathbf{k}\right| \geq 2\sqrt{2}/3\ \pi$. Nevertheless, it must be observed from the previous considerations that the micropolar model provides good simulations of the acoustic behaviour of the discrete model in the smaller domain $-\pi/2 \leq k_i l \leq \pi/2$, $i = 1, 2$, shown in the domain of positivity in Figure 16.b. Based on this consideration and regarding the considered case, one may derive that also for this case the Legendre-Hadamard condition is less strict than the validity of the simulation of the discrete model.

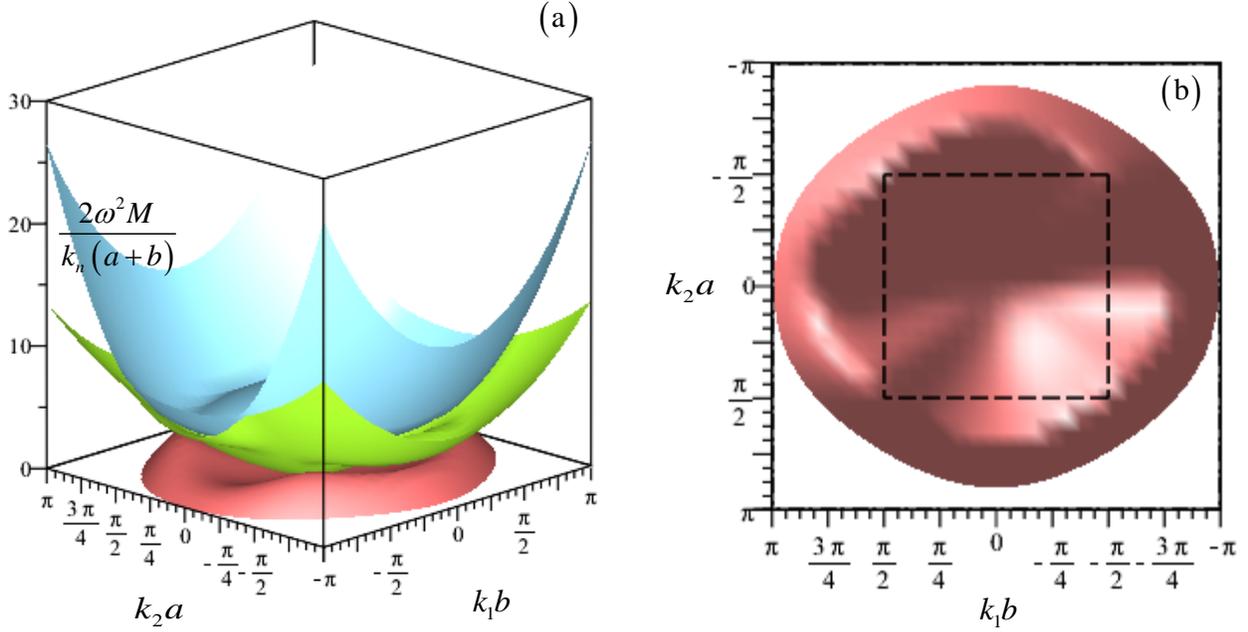

Figure 16: Dispersive surfaces in terms of $2\omega^2 M/k_n(a+b)$ in the Brillouin zone obtained by micropolar continuum model with constitutive tensor $\mathbf{E}_s$ for running bond tessellation ($b/a = 2$, $k_t/k_n = 2/5$). (a) Positive Floquet-Bloch spectrum; (b) Domain of positivity of the first acoustic surface and square subdomain of good accuracy of the model.

## 6. Conclusions

Dispersive waves in two-dimensional blocky materials with periodic microstructure made up of equal rigid units having polygonal centro-symmetric shape with mass and gyroscopic inertia, connected each other through homogeneous linear interfaces, have been analysed. The acoustic behavior of the resulting discrete Lagrangian model has been obtained through a Floquet-Bloch approach. From the resulting eigenproblem derived by the Euler-Lagrange equations for harmonic wave propagation, two acoustic branches and an optical branch are obtained in the frequency spectrum. Moreover, it is shown that the optical branch departs from a critical point with vanishing group velocity in the long wavelength limit. Afterwards, the Lagrangian model has been approximated by a micropolar continuum model following a continualization of the discrete equations of motion, that is based on a downscaling law in which the generalized displacements of the blocks are represented with a second-order Taylor expansion of the generalized macro-displacement field. Moreover, the constitutive equations of the equivalent micropolar continuum have been obtained in a general form with strain and curvature uncoupled because of the assumption of centro-symmetric blocks. The fourth-order elasticity tensor associated to the micropolar strains, that is endowed with major symmetry, turns out to be positive defined. Conversely, the positive definiteness of the second-order symmetric tensor associated to the curvature vector is not guaranteed and depends both on the ratio between the tangent and normal local stiffness and on the block shape.

To verify this outcome, an alternative homogenization has been carried out based on an extended Hamiltonian derivation of the equations of motion for the equivalent continuum. A proper representation of the elastic potential energy has been assumed, that is related to the Hill-Mandel macro homogeneity condition, where a second order expansion of the generalized displacement field is taken into account following an approach suggested by Bazant and Christensen, 1972, in treating rectangular frames. The same equation of motion and the constitutive tensors have been obtained by the two homogenization approaches. Moreover, it has been shown that the hermitian matrix governing the eigenproblem of harmonic wave propagation in the micropolar model is exact up to the second order in the norm of the wave vector with respect to the same matrix from the discrete model. Conversely, if only the first order expansion of the rotational field is retained, when applying the macro-homogeneity condition a remarkably different outcome is achieved, namely a positive defined second order tensor associated to the curvatures is obtained.

To appreciate the acoustic behavior of some relevant blocky materials and to understand the reliability and the validity limits of the micropolar continuum model, some blocky patterns have been analysed: rhombic and hexagonal tilings and running bond masonry. For each blocky assemblage, the dispersion function from the lattice model has been obtained in the reduced Brilluoin zone and compared, in a smaller homothetic region, with the corresponding one from the micropolar continuum model. The rhombic pattern has been analyzed by varying the constitutive parameter, namely the ratio between the tangential and normal local stiffness of the interface. Although the simplicity of the assumed discrete model, rather complex dispersive functions have been obtained and, due to the interaction between the optical and the acoustic branches, no band gap has been observed. Moreover, the micropolar model has been shown to accurately simulate the dispersive functions by the discrete model for wavelengths longer than four times the block size, while rough simulations of the optical branch have been obtained through the continuum model characterized with a positive definite second order tensor. The hexagonal pattern has shown interesting features depending on the ratio between the tangential to the normal stiffness of the interfaces. The micropolar model has turned out very efficient in simulating the dispersive function by the Lagrangian model, with particular reference to the case of long waves where the optical curve of both models shows a downward concavity. Conversely, the micropolar model with positive defined second order tensor has shown a positive concavity of the optical branch, a feature that was criticized by Merkel *et al.*, 2011, when referring to the case of periodic circles, that is a limit case of hexagons with punctual contacts. The running bond masonry or nacreous material has been analysed and a complex band structures without the appearance of band gaps has been obtained. Also in this case, the micropolar model provides a good simulation of the dispersive function from the Lagrangian model, unlike the case of the micropolar model with positive defined second order tensor. From the results obtained in the examples, it appears that the obtained micropolar model turns out to be particularly accurate to describe dispersive functions for wavelengths greater than 3-4 times the characteristic dimension of the block.

Finally, in consideration that the positive definiteness of the second order elastic tensor of the micropolar model is not guaranteed, the hyperbolicity of the equation of motion has been investigated by considering the Legendre–Hadamard ellipticity conditions requiring real values for the wave velocity. Accordingly, since unconditional hyperbolicity cannot be ensured in general, for all the considered examples, the positivity of the square of the dispersion functions has been verified in a homothetic sub-region of the Brillouin zone where a good accuracy of the micropolar model is observed in evaluating the dispersion functions.